%% file: main.tex
\PassOptionsToPackage{table,xcdraw}{xcolor}
\documentclass[a4paper,10pt,twocolumn]{article}

\usepackage{cite}

\usepackage{bmpsize}

\usepackage[colorlinks=true,urlcolor=black]{hyperref}

\def\BibTeX{{\rm B\kern-.05em{\sc i\kern-.025em b}\kern-.08em
T\kern-.1667em\lower.7ex\hbox{E}\kern-.125emX}}

\usepackage{adjustbox}

\usepackage{authblk}

\usepackage{abstract}

\usepackage{geometry}

\usepackage{amsmath,amssymb,amsfonts}

\usepackage{booktabs}

\usepackage{soul}

\usepackage{algorithm}

\usepackage{algorithmicx}

\usepackage[noend]{algpseudocode}

\usepackage{enumitem}

\usepackage{tikz}

\usepackage{graphicx}

\graphicspath{{./images/}}

\usepackage{float}

\usepackage{comment}

\usepackage{subcaption}

\usepackage{makecell}

\usepackage{multirow}

\usepackage{booktabs}

\usepackage{bm}

\usepackage{textcomp}

\usepackage[breakable]{tcolorbox}

\usepackage{xspace}

\usepackage{pifont}

\usepackage{listings}

\newcommand{\mQuery}{f_q}

\newcommand{\query}{$\mQuery$\xspace}

\newcommand{\mPool}{P}

\newcommand{\topk}{top-$k$\xspace}

\newcommand{\mV}{V}

\newcommand{\pool}{$\mPool$\xspace}

\newcommand{\dnn}{DNN\xspace}

\newcommand{\rnn}{RNN\xspace}

\newcommand{\gnn}{GNN\xspace}

\newcommand{\bfs}{BFS\xspace}

\newcommand{\ndcg}{nDCG\xspace}

\newcommand{\recall}{Recall\xspace}

\newcommand{\safe}{SAFE\xspace}

\newcommand{\gemini}{Gemini\xspace}

\newcommand{\jtrans}{jTrans\xspace}

\newcommand{\trex}{Trex\xspace}

\newcommand{\palmtree}{PalmTree\xspace}

\newcommand{\binbert}{BinBERT\xspace}

\newcommand{\clapbfs}{CLAP\xspace}

\newcommand{\reranker}{re-ranker\xspace}

\newcommand{\resafe}{ReSIM\xspace}

\newcommand{\redeep}{DEEP\xspace}

\newcommand{\bincorp}{BinCorp\xspace}

\newcommand{\multicomp}{MultiComp\xspace}

\newcommand{\binpool}{BinPool\xspace}

\newcommand{\ensemble}{\textbf{ENSEMBLE}\xspace}

\newcommand{\cmark}{{\color{green!60!black}\ding{51}}}

\newcommand{\xmark}{{\color{red}\ding{55}}}

\newcommand{\RowW}[9]{%
& #1 & #2 & #3 & #4 & #5 & #6 & #7 & #8 & #9
}

\newcommand{\RowWend}[2]{%
& #1 & #2\\
}

\newcommand{\RowG}[9]{%
& \cellcolor{gray!10}#1 & \cellcolor{gray!10}#2 & \cellcolor{gray!10}#3 &
\cellcolor{gray!10}#4 & \cellcolor{gray!10}#5 &
\cellcolor{gray!10}#6 & \cellcolor{gray!10}#7 &
\cellcolor{gray!10}#8 & \cellcolor{gray!10}#9
}

\newcommand{\RowGend}[2]{%
& \cellcolor{gray!10}#1 & \cellcolor{gray!10}#2\\
}

\colorlet{DeltaPos}{green!55!black}

\definecolor{lightpurple}{RGB}{180,145,255}

\definecolor{mygreen}{RGB}{112,191,73}

\definecolor{Gainsboro}{rgb}{0.86, 0.86, 0.86}

\newtcolorbox{mybox}{
arc=0pt,
boxrule=0pt,
colback=Gainsboro,
width=\columnwidth,
colupper=black
}

\definecolor{babyblueeyes}{rgb}{0.63, 0.79, 0.95}

\newtcolorbox{mybox2}{
arc=0pt,
boxrule=0pt,
breakable,
colback=babyblueeyes,
width=\columnwidth,
colupper=black
}

\tcbset{
cuteanswer/.style={
colback=blue!3!white,
colframe=blue!50!black,
fonttitle=\bfseries,
title=Answer,
boxrule=0.8pt,
arc=4mm,
left=2mm,
right=2mm,
top=1mm,
bottom=1mm,
colbacktitle=blue!15!white,
coltitle=black,
}
}

\tcbset{
rqbox/.style={
colback=teal!3!white,
colframe=teal!50!black,
fonttitle=\bfseries,
title=Research Questions,
boxrule=0.8pt,
arc=4mm,
left=2mm,
right=2mm,
top=1mm,
bottom=1mm,
colbacktitle=teal!10!white,
coltitle=black,
}
}

\definecolor{backcolour}{rgb}{0.95,0.95,0.92}
\definecolor{codegreen}{rgb}{0,0.6,0}
\definecolor{codegray}{rgb}{0.5,0.5,0.5}
\definecolor{codepurple}{rgb}{0.58,0,0.82}
\definecolor{blue}{rgb}{0,0,1}

\lstdefinestyle{mystyle}{
backgroundcolor=\color{backcolour},
commentstyle=\color{codegreen},
numberstyle=\tiny\color{codegray},
stringstyle=\color{codepurple},
basicstyle=\ttfamily\scriptsize,
breakatwhitespace=false,
breaklines=true,
captionpos=b,
keepspaces=true,
numbers=left,
numbersep=5pt,
showspaces=false,
showstringspaces=false,
showtabs=false,
tabsize=2,
aboveskip=0pt,
belowskip=0pt
}

\lstdefinelanguage{MyASM}{
keywords=[1]{mov,lea,xor,movdqu,movups},
keywords=[2]{add,sub,inc,dec,lock},
keywords=[3]{jmp,je,jne,jge,jl},
keywords=[4]{call,ret,retn},
keywords=[5]{push,pop},
keywords=[6]{cmp,test},
keywords=[7]{nop},
keywords=[8]{rax,rbx,rcx,rdi,rsi,rsp,rbp,r12,r13,r14,r15},
keywordstyle=[1]\color{blue}\bfseries,
keywordstyle=[2]\color{orange}\bfseries,
keywordstyle=[3]\color{magenta}\bfseries,
keywordstyle=[4]\color{teal}\bfseries,
keywordstyle=[5]\color{violet}\bfseries,
keywordstyle=[6]\color{red}\bfseries,
keywordstyle=[7]\color{gray}\bfseries,
keywordstyle=[8]\color{black},
sensitive=true,
comment=[l]{;},
morecomment=[s]{/*}{*/},
commentstyle=\color{codegreen},
alsoletter={+,-,[,],*}
}

\def\ackname{Acknowledgments}

\title{\bfseries ReSIM: Re-ranking Binary Similarity Embeddings to Improve Function Search Performance}

\author[1]{Gianluca Capozzi}
\author[2]{Anna Paola Giancaspro}
\author[3]{Fabio Petroni}
\author[2]{Leonardo Querzoni}
\author[2]{Giuseppe Antonio Di Luna}

\affil[1]{KASTEL Security Research Labs, Karlsruhe Institute of Technology\\ \texttt{gianluca.capozzi@kit.edu}}
\affil[2]{DIAG, Sapienza University of Rome, Italy\\ \texttt{\{giancaspro,querzoni,diluna\}@diag.uniroma1.it}}
\affil[3]{European
Molecular Biology Laboratory (EMBL) Rome, Italy\\ \texttt{fabio.petroni@embl.it}}

\date{}

\begin{document}

\twocolumn[

\maketitle

\begin{center}

\begin{minipage}{0.95\textwidth}

\begin{abstract}

\noindent
Binary Function Similarity (\bfs), the problem of determining whether two binary functions originate from the same source code, has been extensively studied in recent research across security, software engineering, and machine learning communities. This interest arises from its central role in developing vulnerability detection systems, copyright infringement analysis, and malware phylogeny tools. Nearly all binary function similarity systems embed assembly functions into real-valued vectors, where similar functions map to points that lie close to each other in the metric space. These embeddings enable function search: a query function is embedded and compared against a database of candidate embeddings to retrieve the most similar matches.

Despite their effectiveness, such systems rely on bi-encoder architectures that embed functions independently, limiting their ability to capture cross-function relationships and similarities. To address this limitation, we introduce \resafe, a novel and enhanced function search system that complements embedding-based search with a neural re-ranker. Unlike traditional embedding models, our reranking module jointly processes query-candidate pairs to compute ranking scores based on their mutual representation, allowing for more accurate similarity assessment. By re-ranking the top results from embedding-based retrieval, \resafe leverages fine-grained relation information that bi-encoders cannot capture.

We evaluate \resafe across seven embedding models on two benchmark datasets, demonstrating consistent improvements in search effectiveness, with average gains of 21.7\% in terms of nDCG and 27.8\% in terms of Recall.

\end{abstract}

\end{minipage}

\end{center}

\vspace{1.5em}

]

\input{sections/introduction}

\input{sections/background}

\input{sections/methodology}

\input{sections/evaluation}

\input{sections/related}

\input{sections/conclusions}

\section*{\ackname} 
This work was supported by the Italian MUR National Recovery and Resilience Plan, funded by the European Union – NextGenerationEU, through the project SERICS (PE00000014). It was also supported by the Agenzia per la Cybersicurezza Nazionale under the 2024–2025 funding programme for the promotion of XL cycle PhD research in cybersecurity - B83C24005660005. The views and opinions expressed are those of the authors and do not necessarily reflect those of the funding institutions.
\bibliographystyle{plain}

\bibliography{bibliography}

\clearpage

\appendix
\input{sections/appendix}

\end{document}

%% file: sections/introduction.tex

\section{Introduction}\label{sec:intro}

In recent years, research on Binary Function Similarity (\bfs) has flourished~\cite{ding2019asm2vec,jTrans-ISSTA22,xu2017neural,massarelli2021function,massarelli2019investigating,DBLP:conf/icml/LiGDVK19,marcelli2022machine,artuso2022binbert,DBLP:journals/tse/PeiXYJR23}.
The \bfs problem consists of assigning a similarity score to two binary functions that is maximized when the functions originate from the same source code (even if compiled with different compilers or optimization levels) and minimized when they are unrelated. 

Most state-of-the-art approaches address \bfs through embedding-based architectures~\cite{jTrans-ISSTA22, artuso2022binbert,DBLP:journals/tse/PeiXYJR23}. Specifically, a Deep Neural Network (\dnn) encodes each function into a fixed-dimensional vector (i.e., the embedding) intended to capture its semantics, and the similarity between two embedded functions is efficiently computed in the embedding space (e.g., using cosine similarity or another distance metric), with higher values assigned to functions stemming from the same source code. This approach is known in information retrieval as the bi-encoder paradigm~\cite{reimers2019sentence,karpukhin2020dense} and enables scalable function search. Specifically, given a large pool of functions (e.g., functions affected by known CVEs), their embeddings can be precomputed, indexed, and queried efficiently through maximum inner product search (MIPS), typically implemented via matrix multiplication, to return the top-$k$ most similar functions to a certain query. In the case of a vulnerability detection system, the retrieved functions correspond to those most likely to contain the same vulnerability as the query function. Embedding-based function search has therefore become the de facto evaluation protocol for \bfs systems~\cite{jTrans-ISSTA22, artuso2022binbert} and underpins a broad range of security applications, including code clone detection, vulnerability detection, copyright infringement analysis, malware analysis and phylogeny, and reverse engineering support.

However, this approach presents a significant limitation. Bi-encoders compute embeddings for the query and each candidate independently, which prevents the model from learning cross-function relationships. In practice, the embedding must condense all potentially relevant information, without knowing which embeddings it will later be compared against. This is particularly limiting in the function search scenario, where the top-ranked results may appear very similar at the binary level (e.g., due to compiler artifacts, common library usages) but originate from different source code. As a result, embedding models may retrieve plausible candidates yet misorder them at the very top positions, degrading retrieval metrics.

In this paper, we address this gap by introducing a re-ranking approach to complement embedding-based function search. First, an embedding model efficiently retrieves the top-$w$ candidates; then, a neural re-ranker reorders these candidates using a joint query-candidate representation to model cross-function interactions and returns the top-$k$ results.

The inspiration for using a \reranker comes from the information retrieval literature. The field of information retrieval has long employed multi-stage architectures to balance efficiency and accuracy \cite{wang2011cascade, nogueira2019multi}.
In the first phase, a recall-oriented candidate retrieval model efficiently identifies potentially relevant items from a large corpus. Modern systems leverage bi-encoder models that independently encode queries and documents into dense vectors. These embeddings enable efficient similarity search via MIPS, but sacrifice fine-grained semantic interactions since queries and documents interact only through vector similarity.
In the second phase, a precision-oriented re-ranking model refines these candidates using more sophisticated scoring. This phase employs cross-encoder architectures that jointly process query-document pairs, enabling deeper semantic interactions \cite{nogueira2019passage,lin2022pretrained}. While computationally expensive, cross-encoders improve retrieval performance and have become standard in Retrieval-Augmented Generation (RAG) systems \cite{lewis2020retrieval,petroni2023improving}.

In this paper, we propose \textbf{\resafe}, a novel two-stage function search pipeline that leverages a cross-encoder re-ranking model to augment embedding-based function search. First, an embedding-based system efficiently retrieves via MIPS the top-$w$ candidates from an indexed database; then, a \reranker jointly processes the query-candidate pairs to re-rank these $w$ functions and return the top-$k$ results, with $k \leq w$. Importantly, \resafe is agnostic to the underlying embedding model.

Experiments show that \resafe significantly improves the \ndcg@$k$ (normalized Discounted Cumulative Gain, a standard metric for search engines~\cite{DBLP:journals/tois/JarvelinK02, DBLP:conf/emnlp/NogueiraJPL20}) and the \recall@$k$ of the considered embedding models. 

\subsection{Contributions}
This paper proposes the following contributions:

\begin{itemize}[noitemsep,topsep=0pt,parsep=0pt,partopsep=0pt]
    \item We train and publicly release the \resafe pipeline for x86-64 assembly functions. Our \reranker is based on the \textbf{DeepSeek-R1-Qwen3-8B}~\cite{guo2025deepseek} model.
    \item We evaluate \resafe by constructing several function search systems, each using a different \bfs embedding model. Specifically, we consider seven embedding models (\gemini~\cite{xu2017neural}, \safe~\cite{massarelli2021function}, \jtrans~\cite{jTrans-ISSTA22}, \clapbfs~\cite{DBLP:conf/issta/WangGZSSZZS0X24}, \binbert~\cite{artuso2022binbert}, \trex~\cite{DBLP:journals/tse/PeiXYJR23}, and \palmtree~\cite{10.1145/3460120.3484587}) chosen for their strong performance and their diversity in terms on DNN architectures and training methodologies.
    \item We demonstrate that, for $k \in \{5, 10, 15, 20, 25, 30\}$ and across two datasets, \resafe consistently improves all evaluated systems (average \recall from 0.59 to 0.72; average \ndcg from 0.69 to 0.82). Moreover, we analyze the impact of the window size $w$ on the computational efficiency and the performance of \resafe.
    \item We show that \resafe is able to ensemble different embedding models, leading to a further 3\% \recall increase on a dataset composed of several toolchains.
    \item We show that there is an important transfer of knowledge from the pre-training of \textbf{DeepSeek-R1-Qwen3-8B}~\cite{guo2025deepseek}, even if it was not tailored on assembly language.
    \item We evaluate \resafe on the vulnerability detection task, considering a large benchmark of CVEs.
    \item We release our fine-tuning and evaluation code at \url{https://github.com/Sap4Sec/reSIM.git}
  
\end{itemize}

%% file: sections/background.tex

\section{Background}\label{sec:background}

This section provides a general overview of the key concepts of embedding-based binary function similarity.

\subsection{Embedding-based Function Search}\label{sec:fsearch}

\begin{figure}[h!]
    \centering
    \includegraphics[width=\linewidth, trim = 0cm 0cm 0cm 0cm]{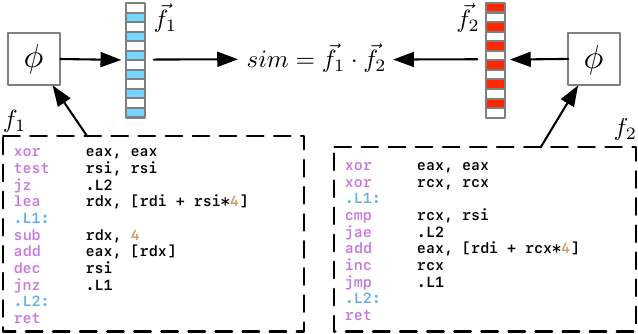}
    \caption{Embedding-based BFS pipeline. The functions $f_1$ and $f_2$ are processed in isolation by the BFS model $\phi$ to produce their embedding representations $\vec{f_1}$ and $\vec{f_2}$, which are then compared using cosine similarity to get the final score.}
    \label{img:biencoder}
\end{figure}

Binary Function Similarity (\bfs) consists of checking whether two binary functions have been compiled from the same source code~\cite{massarelli2021function}. If this condition holds, the two functions are considered to be \textit{semantically similar}.
A {\em \bfs embedding model} is a deep neural network (\dnn) $\phi$ that takes as input an assembly function $f$ and returns a vector $\vec{f} \in \mathbb{R}^{n}$. Given two functions $f_1$ and $f_2$ with embeddings $\vec{f_1} = \phi(f_1)$ and $\vec{f_2} = \phi(f_2)$, their similarity score is measured via cosine similarity $sim=cos(\vec{f}_{1},\vec{f}_2)$. This process is visible in Figure~\ref{img:biencoder}. Ideally, $sim$ should be $1$ when $f_1$ and $f_2$ originate from the same source code, although they may have been compiled with different compilers/optimization levels, and $0$ otherwise. 

Within this context, the \textbf{function search} problem aims to identify the \topk\ functions within a pool that are most similar to a given query. Formally, given a query function \query, a pool of binary functions \pool, and a set $\mV \subseteq \mPool$ of variants of \query (i.e., functions semantically equivalent to \query), we denote by $\mathrm{top}-k(\mQuery,\mPool)$ the set of \topk\ functions returned by the search (ranked by similarity).

In a perfect function search system, every element of $\mV$ appears in the first $|V|$ positions of $\mathrm{top}-k(\mQuery,\mPool)$. That is, an effective function search system maximizes the overlap \(|\mV \cap \mathrm{top}\text{-}k(\mQuery,\mPool)|\). Note that this is only possible when $k \ge |\mV|$; when $k < |\mV|$, the requirement is infeasible. In any case, even when $k < |\mV|$, we can evaluate system performance with standard retrieval metrics. The \recall at rank $k$ is defined as: 
\begin{equation}
\text{Recall@}k(\mQuery) = \frac{|\mV \cap \mathrm{top}-k(\mQuery,\mPool)|}{|\mV|},
\end{equation}
and the normalized discounted cumulative gain at rank $k$ (nDCG@$k$) defined on the ordered list of top-$k$ retrieved results 
$[f_1, f_2, \ldots, f_k]$ as:
\begin{equation}
\text{nDCG@}k(f_q) = 
\frac{
    \sum_{i=1}^k \frac{S(f_i, f_q)}{\log(1 + i)}
}{
    \sum_{i=1}^{min(k, |V|)} \frac{1}{\log(1 + i)}
}.
\label{eq:ndcg}
\end{equation}

Here, $S(f_i, f_q)$ is an indicator function equal to $1$ if $f_i$ is similar to the query function $f_q$, and $0$ otherwise. 
The denominator corresponds to the score of a perfect ranking, while the numerator represents the score achieved by the system under evaluation. 
The \ndcg ranges between $0$ and $1$, and accounts for the ordering of items in the top-$k$ results, rewarding rankings in which relevant items appear earlier.

As an example, consider two result lists for the same query: $(s, s, d, d)$ and $(d, d, s, s)$, 
where $s$ indicates that the corresponding position is occupied by a similar function and $d$ otherwise. 
Although these two results have the same \recall, \ndcg assigns a higher score to the first ranking, since the relevant items appear in earlier positions.

\bfs embedding systems can be adapted to the function search task. Concretely, this is done by transforming the pool $P$ into a set of vectors by applying $\phi$ to each function in $P$, thus obtaining ${\cal P}$. Given a query function $f_q$, its embedding $\vec{f_q} = \phi(f_q)$ is computed and multiplied by each vector in ${\cal P}$, obtaining a similarity score between $f_q$ and each function in $P$. The corresponding functions in $P$ are then ranked according to the aforementioned similarity scores, and the \topk results are returned. Since the vectorization of the pool needs to be computed only once, while the query lookup can be efficiently performed using matrix multiplication followed by sorting, this system design is scalable.

\subsection{BFS Embedding Models}\label{sec:bfs_embd}

BFS embedding models can be classified according to three important dimensions: 
\begin{enumerate}
[noitemsep,topsep=0pt,parsep=0pt,partopsep=0pt]
    \item \textbf{\dnn architecture.} Early work mainly relied on Graph Neural Networks (\gnn)~\cite{xu2017neural, DBLP:conf/icml/LiGDVK19}, and Recurrent Neural Networks (\rnn)~\cite{massarelli2021function}. More recent approaches adopt transformer-based architectures~\cite{DBLP:journals/tse/PeiXYJR23, jTrans-ISSTA22, artuso2022binbert, DBLP:conf/issta/WangGZSSZZS0X24}, to better capture long-range dependencies and handle long instruction sequences.
    \item \textbf{Function representation strategy.} Another important difference is the way they represent assembly functions. \gnn-based methods typically operate on representations derived from the Control Flow Graph (CFG). In some cases~\cite{xu2017neural, DBLP:conf/icml/LiGDVK19}, nodes (i.e., the basic blocks of the CFG) are represented as vectors of manual features (e.g., the number of mathematical operations, the number of strings referenced, ...). Sequence-based models (i.e., \rnn and transformers) consume assembly instruction sequences produced by linear disassembly~\cite{massarelli2021function, jTrans-ISSTA22}, or execution traces~\cite{DBLP:journals/tse/PeiXYJR23, artuso2022binbert}; these instructions are typically preprocessed and normalized to mitigate the out-of-vocabulary (OOV) problem.
    \item \textbf{Training.} \gnn/\rnn-based approaches are commonly trained using a Siamese architecture~\cite{massarelli2021function, DBLP:conf/nips/BromleyGLSS93, reimers2019sentence}. This technique consists of using multiple embedding networks, each producing the embedding of a function, which are then jointly trained by minimizing a loss function (typically, in the form of contrastive or triplet loss). Transformers are typically pre-trained with self-supervised objectives specific to assembly code (e.g., \textit{jump target prediction}~\cite{jTrans-ISSTA22}, \textit{execution language modeling}~\cite{artuso2022binbert}), then fine-tuned for function similarity on pairs or triplets.
\end{enumerate}

These axes have guided our choice of the embedding models to use in our tests, as we will describe in Section~\ref{sec:em}.

\begin{figure*}[h!]
    \centering
    \includegraphics[width=\linewidth, trim = 0cm 0cm 0cm 0cm]{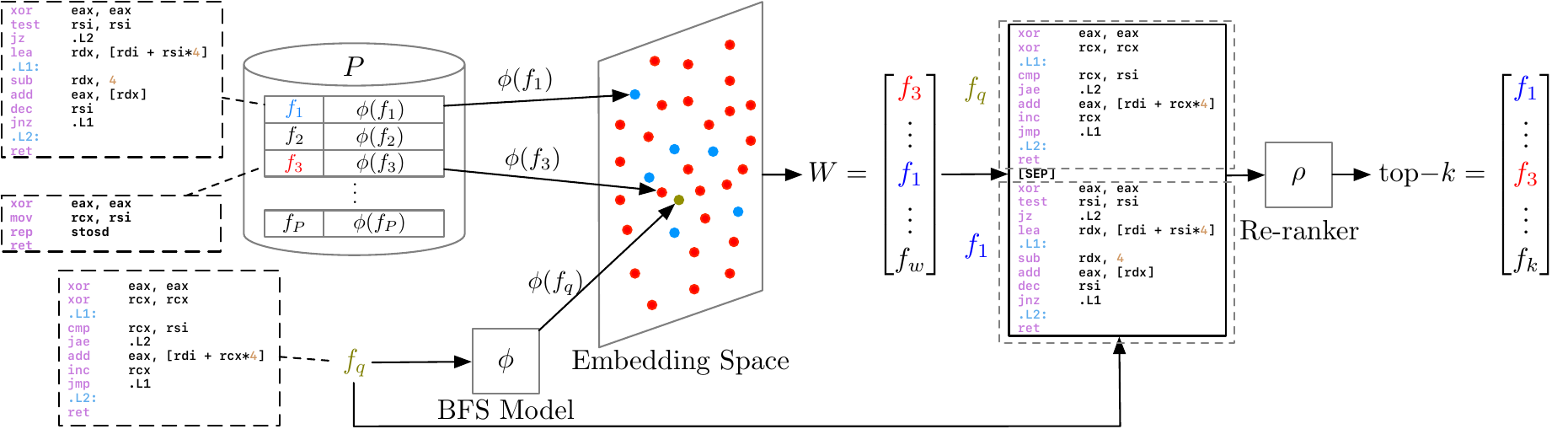}
    \caption{\resafe pipeline. (i) The \bfs\ bi-encoder $\phi$ maps the query function \textcolor{green}{\query} to an embedding $\phi(\mQuery)$. (ii) A similarity measure $sim(\phi(\mQuery),\phi(f))$ is evaluated against the embeddings of all the functions $f \in \mPool$, and the window set $W$ containing the top-$w$ candidates is retrieved; \textcolor{blue}{blue} dots denote embeddings of functions semantically similar to the query, whereas \textcolor{red}{red} dots denote dissimilar ones. (iii) Each $f\in W$ is paired with the query and scored by the re-ranker cross-encoder $\rho(\mQuery,f)\in[0,1]$, which reorders $W$ to produce the final \topk list.}
    \label{img:resim_pipeline}
\end{figure*}

%% file: sections/methodology.tex

\section{\resafe: Re-ranked Function Search}\label{sec:resafe}
Our contribution is \textbf{\resafe}, a two-stage pipeline that enriches the embedding-based function search system described in Section~\ref{sec:fsearch} by integrating a re-ranking neural network. A \reranker is a function $\rho$ that takes as input two assembly functions, $f_1$ and $f_2$, and outputs a ranking score $\rho(f_1, f_2) \in [0,1]$, where values close to $1$ indicate that the two functions are similar, and values close to $0$ indicate dissimilarity. 

Figure~\ref{img:resim_pipeline} illustrates the overall architecture of \textbf{\resafe}. This is based on a \bfs\ embedding model $\phi$ and a re-ranker $\rho$. A query function $f_q$ is transformed into an embedding vector $\vec{f_q} = \phi(f_q)$. The embedding vector is compared with the embedded pool ${\cal P}$ using cosine similarity to obtain a set $W$ of the most similar functions, referred to as the \emph{window set}, of size $w$.
For each pair $(f_q, f_i)$ with $f_i \in W$, the value $\rho(f_q, f_i)$ is computed, yielding an ordered list of rankings $[r_1, r_2, \dots]$. This ranking is used to extract a set of $k$ functions from $W$, which are the \topk results of our function search system.

As an example, let us consider a query \query with $|V|=4$. Assume that the embedding-based function search produces a window $W = [s, d, s, d, s, s]$ of size $6$, where $s$ represents a function similar to the query and $d$ a dissimilar one. In this case, the Recall@$4$ is $2/4$, and the nDCG@$4$ is $0.59$. After applying the re-ranking, we obtain $W' = [s, s, s, d, s, d]$, with a Recall@$4$ of $3/4$ and an nDCG@$4$ of $0.83$.

We stress that the proposed system is general enough that any embedding model $\phi$ can be used. In our experimental evaluation, we will test seven different models, and there are no requirements for using a specific combination of $\phi$ and $\rho$. 

\subsubsection*{Relationship between $W$, $P$, and $k$}
There is a subtle effect determined by the relationship among the size of the window $W$, the size of the pool $P$, and the value of $k$. If $|P| = |W|$, the embedding model $\phi$ becomes irrelevant, and the performance is determined solely by $\rho$. This occurs because all the functions in the pool are ordered by $\rho$, and the final ranking is precisely the one from which the \topk\ functions are selected. Conversely, when $|W| = k$, the order produced by $\rho$ does not influence the functions inside the \topk, and therefore cannot improve the system’s recall at $k$, although it may still affect other metrics such as nDCG. 

This relationship also impacts the total running time of a query lookup. When $|W| = |P|$, the re-ranker $\rho$ must be executed once for each function in $P$. Neglecting possible parallelization and assuming an average runtime $T(\rho)$ for $\rho$, the running time of the re-ranking phase is $|P| \, T(\rho)$. When $|W| = k$, the running time is $k \, T(\rho)$, meaning that execution time  decreases as $|W|$ decreases. However, the relationship between the $|W|$, the \recall and \ndcg is not straightforward. 

It is possible that the embedding model $\phi$ and the re-ranker $\rho$ analyze different aspects of assembly functions. For instance, $\phi$ might leverage execution-related information to produce its embeddings, while $\rho$ relies solely on static features. Therefore, it is not clear whether applying $\rho$ to the entire dataset $P$ would yield better performance than applying it only to the window $W$. It could be that $\phi$ acts as an effective pre-filter, removing certain false positives that $\rho$ would be unable to, thereby providing $\rho$ with a refined window $W$ that ultimately boosts the system's performance. Therefore, the impact of the window size on these outcomes will be analyzed in our experimental evaluation.

\subsubsection*{Ensembling embedding models} 
The basic re-ranking architecture can be extended to use an ensemble of embedding models, $\phi_1,\ldots, \phi_t$.   Using the query, we retrieve several candidate windows $W_1, \ldots, W_t$, where each $W_i$ is obtained from the model $\phi_i$. Each candidate window $W_i$ is first re-ranked individually using $\rho$, producing a set of re-ranked lists $W'_1, \ldots, W'_t$. Each function in these lists now has an associated re-ranker score. The final result is constructed by merging all lists and removing duplicates.  The final, de-duplicated list is then sorted by these scores to produce the top-$k$ results. 
The main motivation for using an ensemble of embedding models is to exploit their diversity (see Section \ref{sec:bfs_embd}), under the assumption that different models may perform better on different binary functions. From our experiments we observe that embedding models fare better on dataset generated with toolchains used during training or fine-tuning. 
It is important to note that the similarity scores produced by different embedding models are \textit{not directly comparable}. The embedding model $\phi_X$ might yield an average similarity of $0.8$ to similar pairs, while a model $\phi_Z$ may assign an average similarity of $0.9$ to similar pair (see~\cite{capozzi2023adversarial}). Because of these differences in scale and calibration across models, one cannot simply aggregate raw similarity scores to construct an ensemble, the re-ranking score avoid this problem. 

\subsection{Re-ranker Architecture}

Following the approach in~\cite{nogueira2019passage}, we instantiate $\rho$ as a transformer-based \textbf{cross-encoder} fine-tuned for binary pairwise ranking. Starting from a pretrained model, we attach a task-specific classification head and fine-tune the model end-to-end. For encoder-only (BERT-like~\cite{vaswani2017}) models, the head operates on the hidden state of the first token; for causal decoder-only models (i.e., Llama~\cite{grattafiori2024llama}, Qwen~\cite{yang2025qwen3}), it uses the hidden state of the last non-padding token. The classification head outputs a single logit, which is used as a ranking score. 

\begin{figure}[h!]
    \centering
    \includegraphics[width=\columnwidth, trim = 0cm 0cm 0cm 0cm]{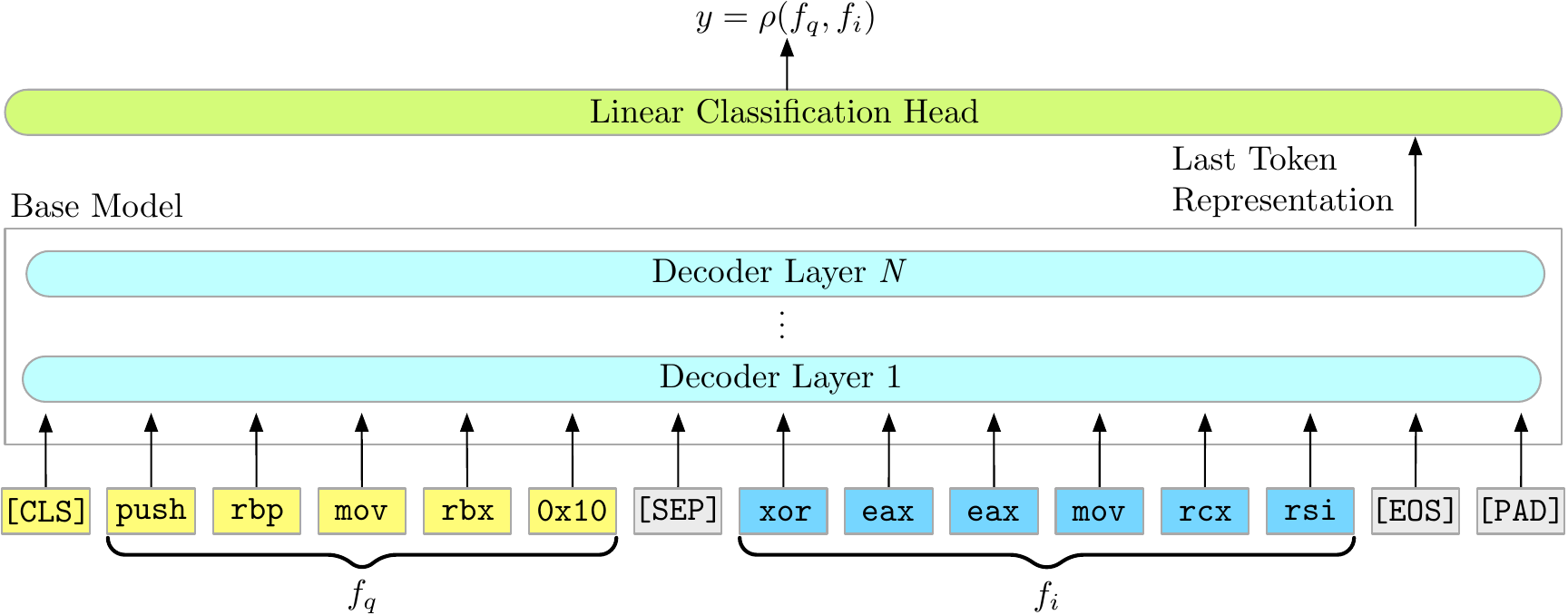}
    \caption{Re-ranking pipeline with a decoder-only transformer $\rho$ (cross-encoder). The two input functions $f_q$ and $f_i$ are tokenized and concatenated with a [\texttt{SEP}] token to form a single sequence processed by the $N$-layer decoder. A classification head reads the hidden state of the last non-padding token and outputs a logit $y$, which is then used to score and re-rank the candidates in $W$.}
    \label{img:reranker}
\end{figure}

Figure~\ref{img:reranker} illustrates the re-ranking process. The \reranker model is provided with the preprocessed linear disassembly of the two functions being compared. Specifically, given a query function $f_q$ and a candidate $f_i \in \mPool$, we concatenate the two sequences with a special separator (e.g., $f_q$ \texttt{[SEP]} $f_i$) to form a single input sequence.

\subsubsection{The DeepSeek-R1-Qwen3-8B Model}
In what follows, we focus on the \textbf{DeepSeek-R1-Qwen3-8B}~\cite{guo2025deepseek} as it is the \reranker architecture we employ for our tests. This is an 8B-parameter dense transformer obtained distilling the reasoning of DeepSeek-R1 into Qwen3-8B~\cite{yang2025qwen3}.

DeepSeek-R1 itself is produced via post-training that combines supervised fine-tuning and reinforcement learning to strengthen reasoning capabilities; DeepSeek-R1-Qwen3-8B inherits these behaviors through distillation into Qwen3-8B. We refer readers to~\cite{guo2025deepseek} for details of the training pipeline.

\subsubsection{Instructions Preprocessing and Tokenization}\label{sec:preprocessing}
To reduce variability and mitigate the OOV problem, we preprocess and normalize the two functions under analysis. We first rebase addresses in the assembly code, then we apply the following normalization steps:
\begin{itemize}[noitemsep,topsep=0pt,parsep=0pt,partopsep=0pt]
    \item Immediate values, memory addresses, and offsets greater than a fixed threshold (5,000) are replaced with the special token \texttt{IMM}.
    \item Jump targets within the function address space are replaced with a relative offset.
    \item User-defined function names are replaced with the special token \texttt{func}.
    \item For calls to \texttt{libc} functions, the target address is replaced with the function name.
\end{itemize}


After concatenating the two normalized sequences using the special token \texttt{[SEP]}, we finally tokenize the resulting sequence using architecture-specific strategies tailored to the re-ranker model (further details are in Section~\ref{sec:impl_details}).

\subsubsection{Re-ranker Finetuning}\label{sec:re_finet}
We fine-tune DeepSeek-R1-Qwen3-8B on the binary pairwise ranking objective: given two binary functions $f_1$ and $f_2$, the model predicts a ranking score $y \in \mathbb{R}$, with higher values assigned to functions compiled from the same source code.

To enhance the \reranker decision boundaries, we incorporate hard negatives during fine-tuning. In particular, different embedding models capture complementary aspects of binary code, so sampling hard negatives from three distinct models allows us to introduce different knowledge aspects into our \reranker model. Thus, for each anchor $f_a$, we construct: a positive ($f_a$, $f^{+}$) pair, where $f^{+}$ is a function semantically equivalent to $f_a$ (i.e., compiled from the same source code~\cite{massarelli2021function}), and three hard negative pairs ($f_a$, $f_{i}^{-}$), with $i \in \{1, 2, 3\}$, where each $f_{i}^{-}$ is sampled from the \topk most similar functions that are not semantically equivalent to $f_a$, according to the \binbert~\cite{artuso2022binbert}, \jtrans~\cite{jTrans-ISSTA22}, and \clapbfs~\cite{DBLP:conf/issta/WangGZSSZZS0X24} models. 

For each triplet ($f_a$, $f^{+}$, $f_{i}^{-}$), we separately compute  $y^{+} = \rho(f_a, f^{+})$ and $y_{i}^{-} = \rho(f_a, f_{i}^{-})$, and then we optimize the margin ranking loss$
    \mathcal{L}(y^{+}, y_{i}^{-}, m) = \mathrm{max}(0, -(y^{+} - y_{i}^{-}) + m)$,
where $m$ is the margin hyperparameter. Intuitively, the loss enforces the positive ranking $y^{+}$ to exceed the negative one $y_{i}^{-}$ by at least $m$, forcing a clear separation between positive and hard negative pairs in the score space.

To reduce trainable parameters and enable efficient fine-tuning of the 8B cross-encoder, we adopt LoRA~\cite{DBLP:journals/corr/abs-2106-09685}, combining it with 4-bit QLoRA quantization~\cite{xu2023qa}. LoRA applies adapters at different projections (i.e., attention-level, MLP) of the base model; each adapter decomposes the corresponding projection matrix into two low-rank adaptations whose output is scaled and added to the frozen projection output. During fine-tuning, only the adapter weights and the classification head are updated while the base model remain fixed. At inference time, the adapter update can be merged into the base weights. Further details about LoRA hyperparameters and target projections are in Section~\ref{sec:impl_details}.
\subsection{BFS Embedding Models}\label{sec:em}
We evaluate \resafe on seven \bfs embedding models that span the main design choices discussed in Section~\ref{sec:bfs_embd} (architecture, function representation, and training). This diversity helps us assess whether re-ranking is consistently beneficial across heterogeneous encoders.

The \textbf{\gemini} model~\cite{xu2017neural} uses a Siamese graph neural network (Structure2Vec~\cite{DBLP:conf/icml/DaiDS16}) over Attributed CFGs, producing function embeddings by aggregating node representations.

The \textbf{\safe} model~\cite{massarelli2021function} embeds linear disassembly with a recurrent architecture trained in a Siamese setup; it builds instruction embeddings (word2vec-inspired~\cite{DBLP:conf/nips/MikolovSCCD13}) and aggregates them with self-attention to obtain a function vector.

The \textbf{\jtrans} model~\cite{jTrans-ISSTA22} adapts BERT to assembly by tokenizing mnemonics and operands and handling jump targets explicitly; it is pretrained with assembly-specific objectives (masked language modeling and jump-target prediction) and then fine-tuned for similarity with a contrastive loss.

The \textbf{\clapbfs} model~\cite{DBLP:conf/issta/WangGZSSZZS0X24} learns assembly embeddings under natural-language supervision by aligning a jump-aware RoBERTa-like encoder~\cite{liu2019roberta} with a text encoder~\cite{sentenceTransf} using contrastive learning (InfoNCE~\cite{oord2018representation}); at inference time, it embeds functions from assembly alone.

The \textbf{\binbert} model~\cite{artuso2022binbert} is an encoder-only transformer pretrained to be execution-aware by leveraging symbolic-execution information alongside assembly; it is then fine-tuned for \bfs in a Siamese/triplet setting.

The \textbf{\palmtree} model~\cite{10.1145/3460120.3484587} is an assembly transformer pretrained with self-supervised objectives on normalized disassembly (e.g., masking and context prediction). Because it outputs instruction-level embeddings, we follow prior work~\cite{artuso2022binbert} and aggregate instructions (via an LSTM) to obtain function-level vectors.

The \textbf{\trex} model~\cite{DBLP:journals/tse/PeiXYJR23} leverages transfer learning with a hierarchical transformer~\cite{DBLP:conf/sigsoft/PeiGBCYWUYRJ21} pretrained on execution traces; for function search it produces embeddings from linear disassembly by pooling final-layer representations and applying a small feed-forward projection.

%% file: sections/evaluation.tex

\section{Datasets and Implementation Details}\label{sec:dataset}
In this section, we describe the datasets used for our experimental evaluation and the implementation details.

\subsection{Fine-tuning Dataset}\label{sec:ft_dataset}
To fine-tune our re-ranker, we use a subset of the \bincorp-26M dataset from~\cite{jTrans-ISSTA22}. According to the authors, this dataset contains 26M functions extracted from 48,130 binaries selected from the official ArchLinux packages and Arch User Repository, compiled for the x86-64 architecture using the gcc compiler with optimization levels ranging from -O0 to -Os. Specifically, we construct a contrastive fine-tuning corpus from the Train split of \bincorp-26M. Following the approach detailed in Section~\ref{sec:re_finet}, we select 833,739 anchor functions and, for each anchor $f_a$, we create three triplets ($f_a$, $f^{+}$, $f_{i}^{-}$), where each anchor has been paired with a random positive sample and three hard negatives. This process yields a total of 2,651,217 triplets.

\subsection{Evaluation Datasets}\label{sec:eval_dataset}
To precisely identify the performance increase attributable to the re-ranker while avoiding confounding factors, we use several distinct datasets for testing. Specifically, we use:

 \noindent {\bf \bincorp Dataset.} From the Test split of \bincorp-26M~\cite{jTrans-ISSTA22}, we select 5,000 query functions. For each query, we consider its set of five semantically similar functions present in the corpus (including the query itself), yielding an evaluation pool of 25,000 functions (see Section~\ref{sec:ft_dataset} for further details). 
 
\noindent {\bf \multicomp Dataset.} This benchmark is a subset of SimTestData~\cite{artuso2022binbert}, obtained by compiling for the x86-64 architecture the following open-source projects: putty0.74, ImageMagick-7.0.10-62, sqlite-3.34.0, gmp-6.2.0, zlib1.2.11, nmap-7.80, and libtomcrypt-1.18.2. In contrast to \bincorp, \multicomp exhibits greater heterogeneity: binaries are produced with multiple versions of gcc (versions 5, 7, and 9), clang (versions 3.8, 6.9, and 9), and icc (version 21), using optimization levels ranging from -O0 to -O3. We select 1,000 queries among the available 5,000, yielding a pool of 11,622 functions. On average, each query has 11.7 similar in the pool.

\begin{table*}
\centering
\caption{\ndcg and \recall for the considered \bfs models using \redeep as reranker ($w=200$, $k \in \{5,10,15,20,25,30\}$), evaluated on pools from \bincorp and \multicomp. Best performances per dataset shown in \textbf{bold}.}
\scriptsize
\label{tab:RQ1}
\begin{tabular}{l|l|l|rrrrrr|rrrrrr}
\cmidrule[\heavyrulewidth]{4-15}
\multicolumn{3}{l}{} & \multicolumn{6}{c|}{\textbf{nDCG}} & \multicolumn{6}{c}{\textbf{Recall}} \\
\midrule
\textbf{Dataset} & \textbf{\bfs model} & \textbf{\reranker} & \textbf{@5} & \textbf{@10} & \textbf{@15} & \textbf{@20} & \textbf{@25} & \textbf{@30} & \textbf{@5} & \textbf{@10} & \textbf{@15} & \textbf{@20} & \textbf{@25} & \textbf{@30}\\
\cmidrule{1-15}
\multirow{14}{*}{\bincorp} & \multirow{2}{*}{\gemini}  & \xmark  & 0.40 & 0.41 & 0.42 & 0.42 & 0.43 & 0.43 & 0.27 & 0.30 & 0.31 & 0.32 & 0.33 & 0.34 \\
 &  & \cmark  & 0.57 & 0.57 & 0.57 & 0.57 & 0.57 & 0.57 & 0.45 & 0.45 & 0.46 & 0.46 & 0.46 & 0.46 \\
\cmidrule{2-15}
 & \multirow{2}{*}{\safe}  & \xmark  & 0.54 & 0.57 & 0.58 & 0.59 & 0.60 & 0.61 & 0.42 & 0.48 & 0.51 & 0.53 & 0.55 & 0.57 \\
 &  & \cmark  & 0.77 & 0.78 & 0.78 & 0.78 & 0.78 & 0.78 & 0.69 & 0.71 & 0.71 & 0.71 & 0.72 & 0.72 \\
\cmidrule{2-15}
 & \multirow{2}{*}{\jtrans}  & \xmark  & 0.75 & 0.80 & 0.82 & 0.83 & 0.84 & 0.84 & 0.68 & 0.77 & 0.81 & 0.83 & 0.85 & 0.86 \\
 &  & \cmark  & \textbf{0.91} & \textbf{0.94} & \textbf{0.95} & \textbf{0.95} & \textbf{0.95} & \textbf{0.95} & \textbf{0.88} & \textbf{0.93} & \textbf{0.94} & \textbf{0.95} & \textbf{0.95} & \textbf{0.95} \\
\cmidrule{2-15}
 & \multirow{2}{*}{\clapbfs}  & \xmark  & 0.87 & 0.89 & 0.90 & 0.91 & 0.91 & 0.91 & 0.83 & 0.87 & 0.89 & 0.90 & 0.91 & 0.91 \\
 &  & \cmark  & 0.91 & 0.94 & 0.94 & 0.94 & 0.94 & 0.94 & 0.88 & 0.92 & 0.94 & 0.94 & 0.94 & 0.95 \\
\cmidrule{2-15}
 & \multirow{2}{*}{\binbert}  & \xmark  & 0.74 & 0.78 & 0.79 & 0.80 & 0.80 & 0.81 & 0.66 & 0.73 & 0.76 & 0.78 & 0.79 & 0.80 \\
 &  & \cmark  & 0.88 & 0.90 & 0.90 & 0.91 & 0.91 & 0.91 & 0.84 & 0.87 & 0.88 & 0.88 & 0.88 & 0.89 \\
\cmidrule{2-15}
 & \multirow{2}{*}{\palmtree}  & \xmark  & 0.59 & 0.61 & 0.62 & 0.63 & 0.63 & 0.64 & 0.48 & 0.52 & 0.54 & 0.56 & 0.57 & 0.58 \\
 &  & \cmark  & 0.76 & 0.76 & 0.77 & 0.77 & 0.77 & 0.77 & 0.68 & 0.69 & 0.69 & 0.70 & 0.70 & 0.70 \\
\cmidrule{2-15}
 & \multirow{2}{*}{\trex}  & \xmark  & 0.61 & 0.65 & 0.66 & 0.67 & 0.67 & 0.68 & 0.50 & 0.56 & 0.60 & 0.62 & 0.63 & 0.65 \\
 &  & \cmark  & 0.81 & 0.82 & 0.82 & 0.82 & 0.82 & 0.82 & 0.74 & 0.76 & 0.76 & 0.77 & 0.77 & 0.77 \\
\midrule
\midrule
\cmidrule{1-15}
\multirow{14}{*}{\multicomp} & \multirow{2}{*}{\gemini}  & \xmark  & 0.61 & 0.49 & 0.46 & 0.46 & 0.47 & 0.48 & 0.25 & 0.31 & 0.34 & 0.36 & 0.38 & 0.40 \\
 &  & \cmark  & 0.84 & 0.71 & 0.67 & 0.66 & 0.66 & 0.66 & 0.39 & 0.51 & 0.54 & 0.55 & 0.56 & 0.56 \\
\cmidrule{2-15}
 & \multirow{2}{*}{\safe}  & \xmark  & 0.74 & 0.62 & 0.60 & 0.61 & 0.62 & 0.63 & 0.33 & 0.43 & 0.49 & 0.53 & 0.56 & 0.58 \\
 &  & \cmark  & 0.92 & 0.85 & 0.82 & 0.81 & 0.81 & 0.81 & 0.45 & 0.66 & 0.72 & 0.74 & 0.75 & 0.76 \\
\cmidrule{2-15}
 & \multirow{2}{*}{\jtrans}  & \xmark  & 0.74 & 0.63 & 0.60 & 0.60 & 0.61 & 0.61 & 0.32 & 0.43 & 0.48 & 0.50 & 0.53 & 0.54 \\
 &  & \cmark  & 0.90 & 0.80 & 0.77 & 0.76 & 0.76 & 0.76 & 0.43 & 0.59 & 0.65 & 0.67 & 0.68 & 0.69 \\
\cmidrule{2-15}
 & \multirow{2}{*}{\clapbfs}  & \xmark  & 0.88 & 0.81 & 0.79 & 0.80 & 0.80 & 0.81 & 0.42 & 0.62 & 0.70 & 0.74 & 0.77 & 0.78 \\
 &  & \cmark  & 0.92 & 0.86 & 0.84 & 0.84 & 0.85 & 0.85 & 0.45 & 0.67 & 0.76 & 0.80 & 0.82 & 0.83 \\
\cmidrule{2-15}
 & \multirow{2}{*}{\binbert}  & \xmark  & 0.91 & 0.86 & 0.84 & 0.84 & 0.85 & 0.86 & 0.45 & 0.67 & 0.76 & 0.80 & 0.83 & 0.84 \\
 &  & \cmark  & \textbf{0.94} & \textbf{0.91} & \textbf{0.90} & \textbf{0.90} & \textbf{0.90} & \textbf{0.91} & \textbf{0.47} & \textbf{0.72} & \textbf{0.82} & \textbf{0.87} & \textbf{0.88} & \textbf{0.89} \\
\cmidrule{2-15}
 & \multirow{2}{*}{\palmtree}  & \xmark  & 0.78 & 0.65 & 0.63 & 0.63 & 0.64 & 0.65 & 0.34 & 0.45 & 0.51 & 0.54 & 0.57 & 0.59 \\
 &  & \cmark  & 0.92 & 0.84 & 0.81 & 0.81 & 0.81 & 0.81 & 0.45 & 0.65 & 0.72 & 0.74 & 0.75 & 0.75 \\
\cmidrule{2-15}
 & \multirow{2}{*}{\trex}  & \xmark  & 0.84 & 0.75 & 0.73 & 0.73 & 0.74 & 0.75 & 0.40 & 0.56 & 0.63 & 0.67 & 0.69 & 0.71 \\
 &  & \cmark  & 0.93 & 0.87 & 0.85 & 0.84 & 0.85 & 0.85 & 0.46 & 0.68 & 0.76 & 0.79 & 0.80 & 0.81 \\
\bottomrule
\end{tabular}
\end{table*}

\subsection{Implementation Details}\label{sec:impl_details}

We implement our re-ranker in Python using the HuggingFace library version 4.57.1. Specifically, for our evaluation, we consider the \redeep re-ranking architecture, finetuned from DeepSeek-R1-Qwen3-8B\footnote{\url{https://huggingface.co/deepseek-ai/DeepSeek-R1-0528-Qwen3-8B}}~\cite{guo2025deepseek} following the procedure described in Section~\ref{sec:re_finet}.
Specifically, we fine-tuned this model following the QLoRA approach~\cite{xu2023qa}, which combines 4-bit quantization with LoRA~\cite{DBLP:journals/corr/abs-2106-09685}. Specifically, we set rank $r=16$, scaling $\alpha = 16$, and dropout $p = 0.05$, targeting the attention (q, k, v, o) and MLP projections (gate, up, down) while keeping biases frozen and initializing LoRA weights.

Furthermore, we fine-tuned \redeep for one epoch with a learning rate of $1\times10^{-4}$, a maximum input length of 2,048 tokens, and a per-device batch size of 8. After preprocessing (see Section~\ref{sec:preprocessing}), we employed left-side tokenization (left truncation), prioritizing the tail of each input function. This is motivated by previous observations that many embedding-based models disproportionately focus on function prologues~\cite{capozzi2025lack, DBLP:conf/icse/WongWLW24}.

For all embedding models, we use the authors' official implementations and publicly released checkpoints, without additional retraining, except for \gemini, for which no official implementation is available.

Finally, we disassemble binaries from the \bincorp and \multicomp codebases with Ghidra\footnote{\url{https://github.com/NationalSecurityAgency/ghidra}}.
The only exceptions are the \jtrans and \clapbfs evaluations on \bincorp, where we used the IDA\footnote{\url{https://hex-rays.com/}} disassembly and scripts supplied with those cases (as required by the original setups).

\section{Experimental Evaluation}\label{sec:evaluation}

Our evaluation comprises the following research questions:
\begin{tcolorbox}[rqbox]

\smallskip

\textbf{RQ1.} What are the performance of \resafe across various compilation toolchains?

\smallskip

\textbf{RQ2.} How is performance affected by the window on which \resafe operates?

\smallskip

\textbf{RQ3.} When does ensembling with \resafe provide performance gains? 
\smallskip

\textbf{RQ4.} Is there a transfer of knowledge from the pre-training of \resafe?

\end{tcolorbox}

Our experimental evaluation was conducted on two hardware platforms. The first (S-A6000) is configured with four A600 GPUs, an AMD Ryzen Threadripper PRO 7985WX (64 cores), and Ubuntu 24.04.2 LTS. The second (S-A100) is an NVIDIA DGX A100 system with eight A100 GPUs, an AMD EPYC 7742 (64 cores), and Ubuntu 22.04.5 LTS.

We evaluate the performance on the function search task using the Recall and nDCG described in Section~\ref{sec:fsearch}.

\begin{table*}[t]
\centering
\caption{Improvement (\%) in nDCG and Recall at @5, @10, @20, and @30 after applying \redeep as reranker.}
\label{tab:improvements_summary}
\begin{adjustbox}{max width=\textwidth}
\scriptsize
\begin{tabular}{l|l|rrrr|r||rrrr|r}
\cmidrule[\heavyrulewidth]{3-12}
\multicolumn{2}{l}{} & \multicolumn{5}{c||}{\textbf{nDCG}} & \multicolumn{5}{c}{\textbf{Recall}} \\
\midrule
\textbf{Dataset} & \textbf{BFS Model} & \textbf{@5} & \textbf{@10} & \textbf{@20} & \textbf{@30} & \textbf{AVG} & \textbf{@5} & \textbf{@10} & \textbf{@20} & \textbf{@30} & \textbf{AVG}\\
\midrule

\multirow{8}{*}{\bincorp}%
\RowG{\gemini}{+42.5\%}{+39.0\%}{+35.7\%}{+32.6\%}{+37.4\%}{+66.7\%}{+50.0\%}{+43.8\%}\RowGend{+35.3\%}{+48.9\%}
\RowW{\safe}{+42.6\%}{+36.8\%}{+32.2\%}{+27.9\%}{+34.9\%}{+64.3\%}{+47.9\%}{+34.0\%}\RowWend{+26.3\%}{+43.1\%}
\RowG{\jtrans}{+21.3\%}{+17.5\%}{+14.5\%}{+13.1\%}{+16.6\%}{+29.4\%}{+20.8\%}{+14.5\%}\RowGend{+10.5\%}{+18.8\%}
\RowW{\clapbfs}{+4.6\%}{+5.6\%}{+3.3\%}{+3.3\%}{+4.2\%}{+6.0\%}{+5.7\%}{+4.4\%}\RowWend{+4.4\%}{+5.2\%}
\RowG{\binbert}{+18.9\%}{+15.4\%}{+13.7\%}{+12.3\%}{+15.1\%}{+27.3\%}{+19.2\%}{+12.8\%}\RowGend{+11.2\%}{+17.6\%}
\RowW{\palmtree}{+28.8\%}{+24.6\%}{+22.2\%}{+20.3\%}{+24.0\%}{+41.7\%}{+32.7\%}{+25.0\%}\RowWend{+20.7\%}{+30.0\%}
\RowG{\trex}{+32.8\%}{+26.2\%}{+22.4\%}{+20.6\%}{+25.5\%}{+48.0\%}{+35.7\%}{+24.2\%}\RowGend{+18.5\%}{+31.6\%}
\cmidrule{2-12}

\rowcolor{blue!10}
\cellcolor{white} & \textbf{AVG} & \textbf{+27.4\%} & \textbf{+23.6\%} & \textbf{+20.6\%} & \textbf{+18.6\%} & \textbf{+22.5\%} & \textbf{+40.5\%} & \textbf{+30.3\%} & \textbf{+22.7\%} & \textbf{+18.1\%} & \textbf{+27.9\%} \\
\midrule

\multirow{8}{*}{\multicomp}%
\RowG{\gemini}{+37.7\%}{+44.9\%}{+43.5\%}{+37.5\%}{+40.9\%}{+56.0\%}{+64.5\%}{+52.8\%}\RowGend{+40.0\%}{+53.3\%}
\RowW{\safe}{+24.3\%}{+37.1\%}{+32.8\%}{+28.6\%}{+30.7\%}{+36.4\%}{+53.5\%}{+39.6\%}\RowWend{+31.0\%}{+40.1\%}
\RowG{\jtrans}{+21.6\%}{+27.0\%}{+26.7\%}{+24.6\%}{+25.0\%}{+34.4\%}{+37.2\%}{+34.0\%}\RowGend{+27.8\%}{+33.3\%}
\RowW{\clapbfs}{+4.5\%}{+6.2\%}{+5.0\%}{+4.9\%}{+5.2\%}{+7.1\%}{+8.1\%}{+8.1\%}\RowWend{+6.4\%}{+7.4\%}
\RowG{\binbert}{+3.3\%}{+5.8\%}{+7.1\%}{+5.8\%}{+5.5\%}{+4.4\%}{+7.5\%}{+8.7\%}\RowGend{+6.0\%}{+6.7\%}
\RowW{\palmtree}{+17.9\%}{+29.2\%}{+28.6\%}{+24.6\%}{+25.1\%}{+32.4\%}{+44.4\%}{+37.0\%}\RowWend{+27.1\%}{+35.2\%}
\RowG{\trex}{+10.7\%}{+16.0\%}{+15.1\%}{+13.3\%}{+13.8\%}{+15.0\%}{+21.4\%}{+17.9\%}\RowGend{+14.1\%}{+17.1\%}
\cmidrule{2-12}

\rowcolor{blue!10}
\cellcolor{white} & \textbf{AVG} & \textbf{+17.2\%} & \textbf{+23.7\%} & \textbf{+22.7\%} & \textbf{+19.9\%} & \textbf{+20.9\%} & \textbf{+26.5\%} & \textbf{+33.8\%} & \textbf{+28.3\%} & \textbf{+21.8\%} & \textbf{+27.6\%} \\

\bottomrule
\end{tabular}
\end{adjustbox}
\end{table*}

\subsection{RQ1: Testing Across  Toolchains}\label{sec:rq1}

We evaluate \resafe on seven embedding models (i.e., \gemini, \safe, \jtrans, \clapbfs, \binbert, \palmtree, and \trex), considering the \bincorp and \multicomp datasets described in Section~\ref{sec:dataset}. The rationale for employing multiple datasets is to assess the effectiveness of \resafe while mitigating potential biases that could arise from evaluating on a limited set of compilers or binaries: differently from the \bincorp dataset, whose binaries are compiled using the same toolchain as the fine-tuning corpus,  \multicomp contains binaries produced by toolchains that are not used in the fine-tuning dataset.

Table~\ref{tab:RQ1} presents the \ndcg and \recall scores for all embedding models with and without \redeep re-ranking, evaluated at $k \in \{5, 10, 15, 20, 25, 30\}$ across both datasets using a window size of $w=200$. To provide a clearer overview of the benefits introduced by our re-ranker, Table~\ref{tab:improvements_summary} reports the corresponding percentage improvements for $k \in \{5, 10, 20, 30\}$.

As shown in Table~\ref{tab:improvements_summary}, the degree of improvement varies considerably across \bfs models and datasets. The most substantial gains are observed for older embedding models, namely \gemini and \safe. On \bincorp, \gemini achieves average \ndcg gains of 37.4\% and \recall improvements of 48.9\%, while \safe shows \ndcg gains of 34.9\% and \recall improvements of 43.1\%. Similar trends emerge on \multicomp, with \gemini improving by 40.9\% in \ndcg and 53.3\% in \recall, and \safe by 30.7\% in \ndcg and 40.1\% in \recall.

Modern transformer-based models still show significant improvements, typically in the 4--35\% range on both metrics. 

Specifically, on \bincorp, \trex benefits the most from re-ranking, with average gains of 25.5\% in \ndcg and 31.6\% in \recall. On \multicomp, \palmtree shows the largest improvement, with average increases of 25.1\% in \ndcg and 35.2\% in \recall. Overall, these results indicate that \redeep consistently improves performance across different embedding models.

In the remainder of this section, we focus on the top three re-ranked models (\clapbfs, \jtrans, and \binbert).

\begin{figure}[h!]
    \centering
    \includegraphics[width=\columnwidth, trim = 0cm 0cm 0cm 0cm]{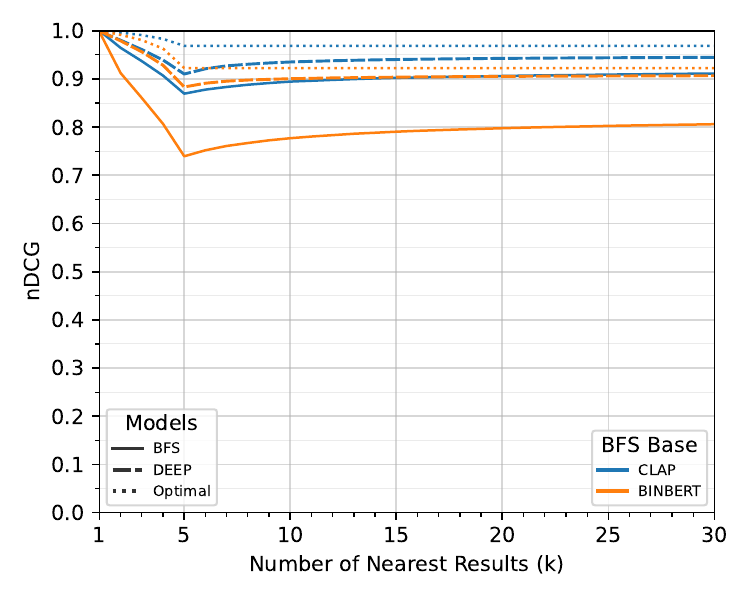}
    \caption{\ndcg for \clapbfs and \binbert with \redeep re-ranking ($w=200$, $k \in [1,30]$), evaluated on a pool of 25,000 functions and 5,000 queries from \bincorp.}
    \label{img:RQ2_ndcg_bincorp}
\end{figure}

\subsubsection{Results on \bincorp.}

\begin{figure}[h!]
    \centering
    \includegraphics[width=\columnwidth, trim = 0cm 0cm 0cm 0cm]{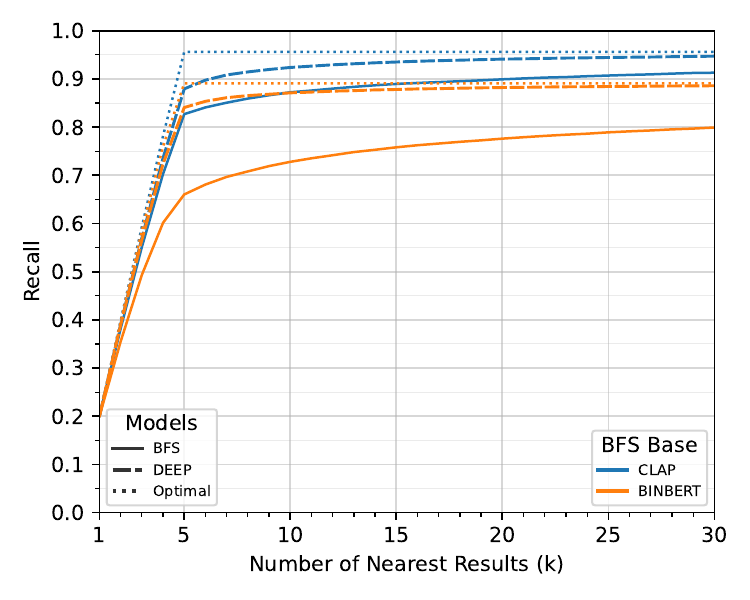}
    \caption{\recall for \clapbfs and \binbert with \redeep re-ranking ($w=200$ and $k \in [1,30]$), evaluated on a pool of 25,000 functions and 5,000 queries from \bincorp.}
    \label{img:RQ2_recall_bincorp}
\end{figure}

Figures~\ref{img:RQ2_ndcg_bincorp} and~\ref{img:RQ2_recall_bincorp} show \ndcg and \recall for \clapbfs and \binbert on \bincorp: solid lines for baseline performance and dashed lines for \redeep. \jtrans results are in Appendix~\ref{sec:app_B} (Figures~\ref{img:RQ2_jtrans_bincorp_1}-\ref{img:RQ2_jtrans_bincorp_2}).

Without re-ranking, \clapbfs dominates across all $k$ values for \ndcg (0.87--0.91) and \recall (0.83--0.91), significantly outperforming \binbert. After applying \redeep, as shown in Figures~\ref{img:RQ2_ndcg_bincorp} and~\ref{img:RQ2_recall_bincorp}, the performance gap between \clapbfs and \binbert substantially narrows, with both models achieving comparable results. From our experiments, \jtrans + \redeep achieves comparable performance (0.94--0.95 \ndcg) and slightly surpasses \clapbfs + \redeep in \recall by 0.01 points from $k \geq 10$. Although only \jtrans and \clapbfs were trained on \bincorp functions (unlike \binbert), \redeep brings all three models to similar performance levels.

\subsubsection{Results on \multicomp.}
\begin{figure}[h!]
    \centering
    \includegraphics[width=\columnwidth, trim = 0cm 0cm 0cm 0cm]{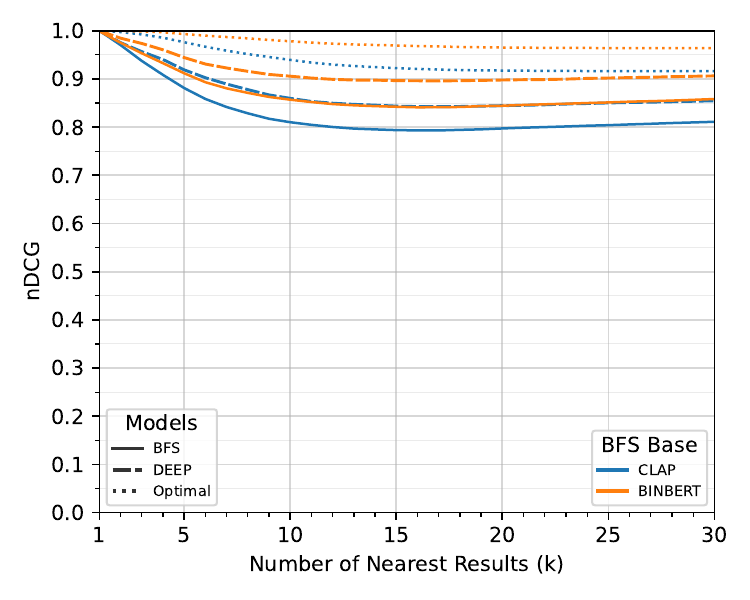}
    \caption{nDCG for \clapbfs and \binbert with \redeep re-ranking (with $w=200$, $k \in [1,30]$), evaluated on a pool of 11,622 functions and 1,000 queries from \multicomp.}
    \label{img:RQ2_ndcg_multicomp}
\end{figure}

\begin{figure}[h!]
    \centering
    \includegraphics[width=\columnwidth, trim = 0cm 0cm 0cm 0cm]{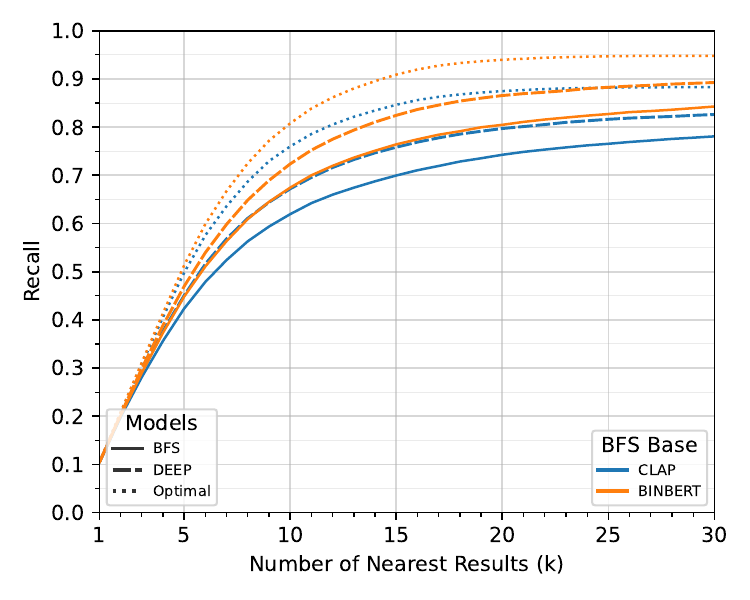}
    \caption{Recall for \clapbfs and \binbert with \redeep re-ranking ($w=200$ and $k \in [1,30]$), evaluated on a pool of 11,622 functions and 1,000 queries from \multicomp.}
    \label{img:RQ2_recall_multicomp}
\end{figure}

Figures~\ref{img:RQ2_ndcg_multicomp} and~\ref{img:RQ2_recall_multicomp} show the results for \clapbfs and \binbert on \multicomp. For completeness, results for \jtrans are shown in Appendix~\ref{sec:app_B} (Figures~\ref{img:RQ2_jtrans_multicomp_1}-\ref{img:RQ2_jtrans_multicomp_2}). We remark that \redeep has not been finetuned on binaries from \multicomp.

Without re-ranking, \binbert is the best performing model with average \ndcg of 0.87 and \recall of 0.68, as expected given its fine-tuning on the same toolchains included in \multicomp. However, \redeep is still able to increase the performance, with \binbert + \redeep being the best-performing model across all $k$ values, with an average \ndcg of 0.91 and \recall of 0.73. For \clapbfs and \jtrans, \redeep improves the baseline average performance by 15.1\% in \ndcg and 20.4\% in \recall. These results highlight \redeep’s strong generalization across unseen compilation toolchains.

\subsubsection{Comparison with Optimal Performance}\label{sec:opt_perf}
In the following, we examine how close \redeep approaches theoretical upper bound. In Figures~\ref{img:RQ2_ndcg_bincorp} and~\ref{img:RQ2_recall_bincorp} (for \bincorp) and Figures~\ref{img:RQ2_ndcg_multicomp} and~\ref{img:RQ2_recall_multicomp} (for \multicomp), we show with the dotted line the performance of the optimal re-ranker.

We can compute the optimal re-ranking for each query using the ground-truth labels in our datasets. Consider a query for which the pool \pool contains five semantically similar (i.e., relevant) functions. If the embedding-based retriever returns only four of these five functions within the current candidate window, the optimal re-ranker would place all four retrieved relevant functions at the top of the ranking.

The gap between the embedding-based retrieval performance and the oracle upper bound indicates that, while bi-encoders identify most relevant functions within \pool, they do not rank them optimally. This gap quantifies the headroom available to re-ranking methods, and the gap between the upper bound and \redeep quantifies the goodness of our approach.

Consider \jtrans on the \bincorp dataset: without re-ranking, it achieves average \ndcg and \recall values of 0.83 and 0.76, respectively, falling short of the optimal 0.98 (\ndcg) and 0.90 (\recall). \redeep improves performance to the near-optimal values of 0.95 for \ndcg and 0.88 for \recall. For \clapbfs, the initial average performance stands at 0.91 (\ndcg) and 0.83 (\recall), compared to optimal values of 0.97 and 0.90. \redeep pushes performance to 0.94 for \ndcg and 0.87 for \recall.
Finally, on the \multicomp dataset, \binbert achieves average \ndcg and \recall values of 0.87 and 0.68, compared to optimal performance of 0.97 and 0.79. Notably, although \redeep’s fine-tuning dataset does not include the compilation toolchain used in \multicomp, it still improves performance to 0.91 in \ndcg and 0.73 in \recall.

\begin{tcolorbox}[cuteanswer,title={Answer to RQ1}]
On \bincorp, the \reranker enhances performance across all models, yielding average gains of 22.5\% in \ndcg and 27.9\% in \recall. The best-performing configuration is \resafe with \jtrans as the embedding model.

On \multicomp, our approach still yields substantial improvements of 20.9\% in \ndcg and 27.6\% in \recall. Here, \resafe with \binbert as the underlying embedding model achieves the highest performance.

Overall, the performance gap between our \redeep and the optimal \reranker is 3.8\% for \ndcg and 4.4\% for \recall.
\end{tcolorbox}

\subsection{RQ2: Impact of the Window $W$}\label{sec:rq2}

We now discuss the impact of the window size $w$ on both the computational efficiency and the performance of \resafe.  
As explained in Section~\ref{sec:resafe}, increasing or decreasing the value of $w$ affects the time required to execute the re-ranker and may also influence Recall and nDCG.  
We restrict this analysis to the top three embedding models identified while answering RQ1: \jtrans, \clapbfs, and \binbert.

\begin{figure}[h!]
    \centering
    \includegraphics[width=\columnwidth, trim = 0cm 0cm 0cm 0cm]{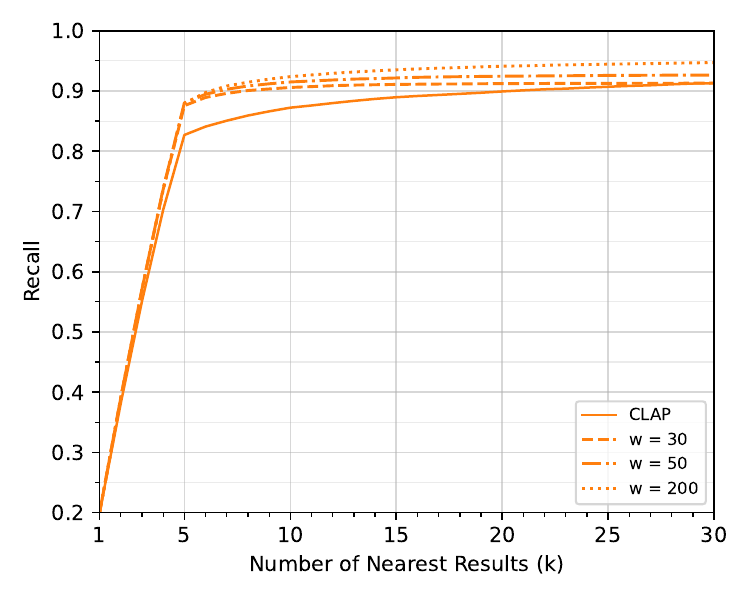}
    \caption{Recall for the \clapbfs model with \redeep re-ranking, ($w \in \{30, 50, 200\}$, $K \in [1,30]$) evaluated on a pool of 25,000 functions and 5,000 queries from \bincorp.}
    \label{img:RQ2_recall_clap}
\end{figure}

\subsubsection{Impact on \ndcg and \recall}\label{sec:wimpact}
As expected, $w$ acts as a parameter that influences the performance of the function search. 

On \bincorp, the re-ranked \clapbfs achieves optimal overall performance at $w=200$. The \ndcg metric remains relatively stable across all window sizes for the considered values of $k$. In contrast, as visible in Figure~\ref{img:RQ2_recall_clap}, \recall exhibits more sensitivity to $w$: the performance is comparable across window sizes for $k \le 10$, but diverges at higher $k$ values, where larger windows provide clear advantages (e.g., while at $k=5$ \recall coincides across all the values of $w$, at $k=30$, \recall increases from 0.91 for $w=30$ to 0.95 for $w=200$). Similar patterns emerge for \clapbfs on \multicomp.

\begin{figure}[h!]
    \centering
    \includegraphics[width=\columnwidth, trim = 0cm 0cm 0cm 0cm]{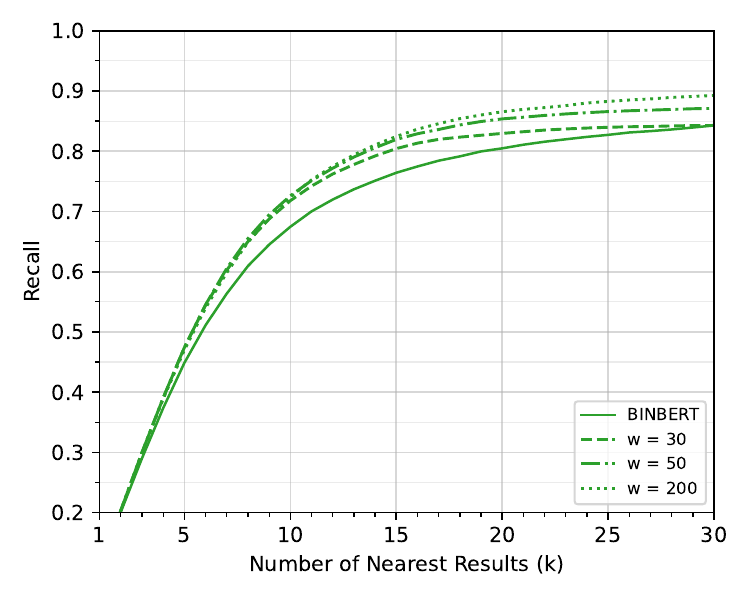}
    \caption{Recall for the \binbert model with \redeep re-ranking ($w \in \{30, 100, 200\}$, $k \in [1,30]$), evaluated on a pool of 11,622 functions and 1,000 queries from \multicomp.}
    \label{img:RQ2_recall_binbert}
\end{figure}

The results for \jtrans and \binbert on \bincorp show that $w=200$ yields the best performance for both models, with both metrics exhibiting greater sensitivity to window size compared to \clapbfs. On this dataset, both models show marked differences even at low $k$ values, with average absolute differences at $k=30$ of 0.06 for \ndcg and 0.09 for \recall. 

On \multicomp, the two models exhibit different behaviors. \jtrans maintains the same performance gaps across all $k$ values seen on \bincorp, whereas, as reported in Figure~\ref{img:RQ2_recall_binbert}, \binbert demonstrates minimal sensitivity for $k \le 10$, with gaps emerging only as $k$ increases.

Full results are reported in Table~\ref{tab:RQ2} (see Appendix~\ref{sec:app_A}).

\subsubsection{Impact on Computational Efficiency}\label{sec:timew} 

Increasing the window size $w$ increases the running time of the re-ranker module. Specifically, the total execution time for processing a query can be decomposed into three components:  
the time $t(\phi_i)$ for the embedding model $\phi_i$ to compute the embedding of the query function,  
the time $t(\mathrm{sim})$ required to obtain the candidate window $W$ based on embedding similarity,  
and the time $t(\rho)$ required by the re-ranker to reorder the window $W$. While $t(\phi_i)$ and $t(\mathrm{sim})$ are effectively independent of $w$, $t(\rho)$ scales with it.

In the following, we evaluate $t(\rho)$ when performing the re-ranking step for a single query on the \texttt{S-A6000} hardware. We highlight that this value is independent of the specific embedding model $\phi_i$, while it strictly depends on $w$. Indeed, $t(\rho)$ scales linearly with the window size $w$ (e.g., halving $w$ from 200 to 100 approximately halves the execution time; the same holds for $w = 50$ and $w = 30$). This observation is consistent with our theoretical analysis in Section~\ref{sec:resafe}. Specifically, $t(\rho)$ is equal to 16.5 seconds at $w=30$, 27.6 seconds at $w=50$, 54.9 seconds at $w=100$, and 109.4 seconds at $k=200$.

Both $t(\phi_i)$ and $t(\mathrm{sim})$ are negligible compared to $t(\rho)$. Specifically, when considering the \jtrans, \clapbfs, and \binbert models, $t(\phi_i)$ exhibits an average of 0.013 seconds, while $t(\mathrm{sim})$ is constant at 0.006 seconds. This is expected because $\mathrm{sim}$ is a vector–matrix multiplication, and the three models produce 768-dimensional embeddings. Therefore, the computational efficiency of \resafe can be analyzed primarily as a function of $w$, while neglecting the other contributions.
We stress that these times are for batch size; during $t(\mathrm{sim})$, we compute the re-ranking score in parallel for all pairs that fit a batch size. The batch size depends on the architecture; in our test, we managed to use 50 as the batch size.

\begin{table*}
\centering
\caption{\ndcg and \recall for the considered \bfs models with \redeep re-ranking ($w=200$, $k \in \{5, 10, 15, 20, 25, 30\}$), evaluated on a pool of 8,441 functions jointly extracted from \bincorp and \multicomp. The \ensemble row shows the performance obtained when ensembling the results of the two best-performing re-ranked models (i.e., \jtrans and \binbert).}
\scriptsize
\label{tab:RQ3}
\begin{tabular}{l|l|rrrrrr|rrrrrr}
\cmidrule[\heavyrulewidth]{3-14}
\multicolumn{2}{l}{} & \multicolumn{6}{c|}{\textbf{nDCG}} & \multicolumn{6}{c}{\textbf{Recall}} \\
\midrule
\textbf{\bfs model} & \textbf{\reranker} & \textbf{@5} & \textbf{@10} & \textbf{@15} & \textbf{@20} & \textbf{@25} & \textbf{@30} & \textbf{@5} & \textbf{@10} & \textbf{@15} & \textbf{@20} & \textbf{@25} & \textbf{@30}\\
\midrule
\multirow{2}{*}{\jtrans} & \xmark & 0.82 & 0.79 & 0.78 & 0.78 & 0.79 & 0.80 & 0.58 & 0.68 & 0.72 & 0.75 & 0.76 & 0.78 \\
& \cmark & 0.93 & 0.90 & 0.89 & 0.89 & 0.89 & 0.89 & 0.68 & 0.80 & 0.84 & 0.86 & 0.86 & 0.87 \\
\midrule
\multirow{2}{*}{\binbert} & \xmark & 0.87 & 0.86 & 0.86 & 0.86 & 0.87 & 0.87 & 0.60 & 0.75 & 0.81 & 0.84 & 0.86 & 0.87 \\
& \cmark & 0.94 & 0.93 & 0.93 & 0.93 & 0.93 & 0.93 & 0.69 & 0.83 & 0.88 & 0.91 & 0.91 & 0.92 \\
\midrule
\multirow{2}{*}{\clapbfs} & \xmark & 0.92 & 0.90 & 0.89 & 0.89 & 0.90 & 0.90 & 0.67 & 0.79 & 0.84 & 0.87 & 0.88 & 0.89 \\
& \cmark & 0.94 & 0.92 & 0.92 & 0.92 & 0.92 & 0.92 & 0.69 & 0.82 & 0.87 & 0.90 & 0.91 & 0.91 \\
\midrule
\midrule
\multicolumn{2}{c|}{\ensemble} & 0.94 & 0.94 & 0.94 & 0.94 & 0.94 & 0.95 & 0.69 & 0.85 & 0.90 & 0.93 & 0.94 & 0.95 \\
\bottomrule
\end{tabular}
\end{table*}

\begin{tcolorbox}[cuteanswer,title={Answer to RQ2}]
The window size $w$ significantly impacts \resafe performance. Larger windows (e.g., $w=200$) consistently yield better results for every value of $k \in \{1, 30\}$, with the performance gains exhibiting variability depending on the considered embedding model.

The re-ranking module dominates the per-query execution time, which increases approximately linearly with $w$. In contrast, the costs of embedding computation and similarity search are negligible. As a result, the end-to-end execution time is under one second without re-ranking, around 16 seconds at $w = 30$, and approximately 109 seconds at $w = 200$.
\end{tcolorbox}

\begin{figure}[h]
    \centering
    \includegraphics[width=\columnwidth, trim = 0cm 0cm 0cm 0cm]{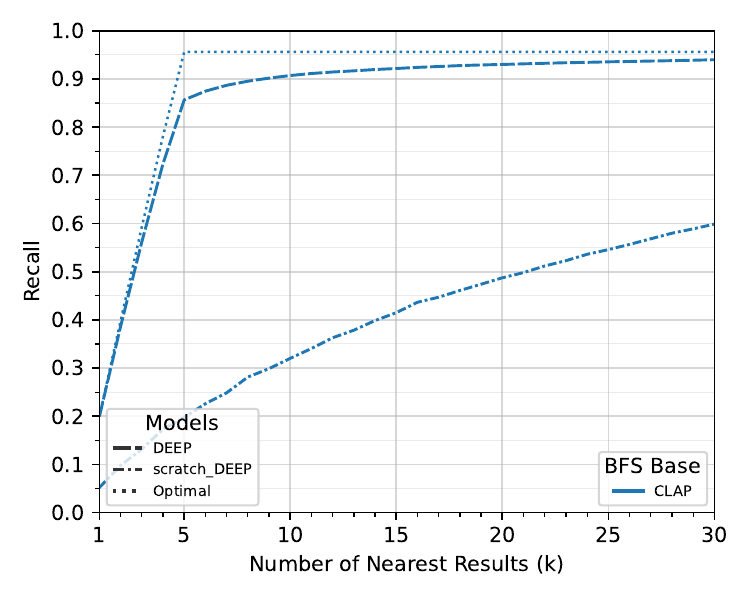}
    \caption{nDCG for the \clapbfs model with \redeep and scratch\_\redeep re-ranking ($w=200$, $K \in [1,30]$), evaluated on a pool of 25,000 functions and 5,000 queries from \bincorp.}
    \label{img:RQ4_impact_nDCG}
\end{figure}

\subsection{RQ3: Ensembling Models with \resafe}\label{sec:rq3}

In this section, we test the ensembling methodology. We ensemble the two best-performing re-ranked embedding models from Section~\ref{sec:rq1}.
This is done as described in Section~\ref{sec:resafe}: we retrieve the top 100 results by applying \resafe to \jtrans and \binbert, then order them by score and remove duplicates.

We evaluate this ensemble strategy on a dataset comprising functions from \bincorp and \multicomp. Specifically, starting from the pools described in Section~\ref{sec:eval_dataset}, we select 1000 queries (500 from \bincorp and 500 from \multicomp), resulting in a final pool of 8,441 functions.

As expected, \redeep enhances the performance of the evaluated \bfs models with respect to both \ndcg and \recall. These improvements are further amplified through ensembling. Compared to the best individual re-ranked model on this pool (i.e., \redeep applied to \clapbfs), our aggregation yields an additional 2\% increase in \ndcg and 3\% in \recall, with the gains being more pronounced for larger values of $k$ ($k \ge 10$). The results are reported in Table~\ref{tab:RQ3}. As we can see, the ensembling strategy outperforms both the non-re-ranked embedding models and the re-ranked ones. This arises from ensembling two embedding models, each of which exhibits state-of-the-art performance on a different toolchain. 

Interestingly, the approach yields gains also on \bincorp (+1\% in \ndcg and +2\% in \recall) and matches the \redeep re-ranked \binbert on \multicomp.  Details are in Appendix~\ref{sec:app_C}.

\begin{tcolorbox}[cuteanswer,title={Answer to RQ3}]
Ensembling embedding models yields a $3\%$ Recall improvement on mixed datasets composed of multiple toolchains, outperforming any single embedding model. We believe this strategy is particularly effective on datasets built from toolchains that were not represented in the training data of an individual embedding model. On datasets composed of toolchains already seen during training, ensembling still improves or matches the performance of the non-ensembled re-ranking.
\end{tcolorbox}

\begin{figure}[h!]
    \centering
    \includegraphics[width=\columnwidth, trim = 0cm 0cm 0cm 0cm]{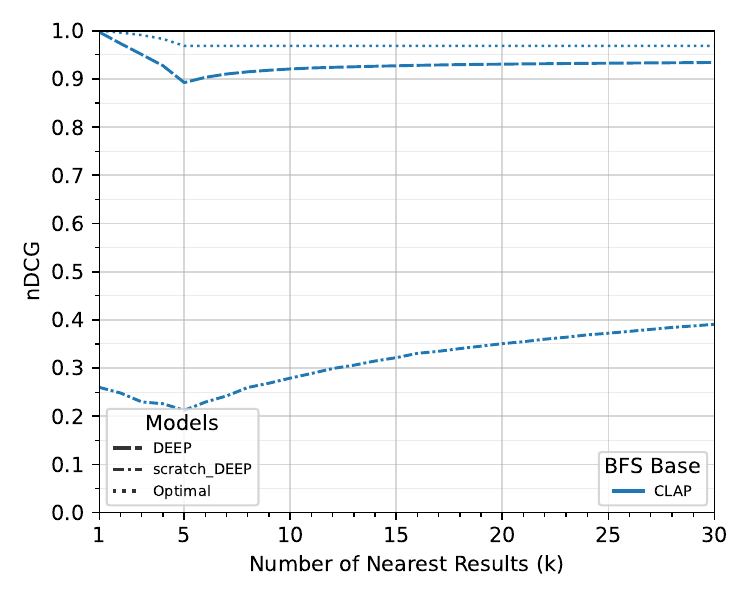}
    \caption{\recall for the \clapbfs model with \redeep and scratch\_\redeep re-ranking, ($w=200$, $K \in [1,30]$), evaluated on a pool of 25,000 functions and 5,000 queries from \bincorp.}
    \label{img:RQ4_impact_recall}    
\end{figure}

\subsection{RQ4: Usefulness of Pre-Training}\label{sec:rq4}

In this section, we investigate the impact of the transformer architecture and of the pre-training data on \resafe. One may wonder whether the pre-training of \redeep is actually useful, as it is not specifically trained on assembly code and one could argue that most of the data it observes is not assembly. We show that this hypothesis is false: pre-training has a major impact on re-ranking performance.

To evaluate the impact of pre-training, we randomize 91.77\% of the weights of \redeep (we randomize all layers but the embedding and normalization one). This effectively obliterates all the knowledge stored in the model. We then fine-tune this ``blank'' model using the QLoRA methodology described in Section~\ref{sec:re_finet}.

To avoid unnecessary energy consumption, we compare the performance of the blank model (scratch\_\redeep) against the pre-trained one (\redeep) at the fifth training checkpoint. Ensuring that both models have seen the same data and training steps. A large performance gap at this early checkpoint is a strong indication that the pre-training data is indeed useful.

We evaluate random\_\redeep and \redeep’s ability to re-rank \clapbfs results on the \bincorp pool from Section~\ref{sec:eval_dataset}. The results in Figure~\ref{img:RQ4_impact_recall} and Figure~\ref{img:RQ4_impact_nDCG} clearly show pre-training’s significant benefit: at the fifth checkpoint, pre-trained \redeep on \clapbfs achieves an average of 0.92 \ndcg and 0.93 \recall, whereas scratch\_\redeep on \clapbfs achieves 0.14 in \ndcg and 0.43 in \recall. 
We stress that QLoRA fine-tuning of \redeep updates only a small subset of parameters (namely, 0.57\% of the total, which accounts for 43,655,168 parameters\footnote{For comparison, a BERT transformer has roughly 110 million parameters}), which further indicates that the pre-training knowledge acquired by \redeep is instrumental.

\begin{tcolorbox}[cuteanswer,title={Answer to RQ4}]
Our experiments show a substantial transfer of information from the pre-trained \redeep. This may appear surprising, since its pre-training data is mostly natural language and high-level code. We speculate that this is because even non-domain-tailored pre-training improves the model's reasoning capabilities. 
\end{tcolorbox}

\begin{figure}[h!]
    \centering
    \includegraphics[width=\columnwidth, trim = 0cm 0cm 0cm 0cm]{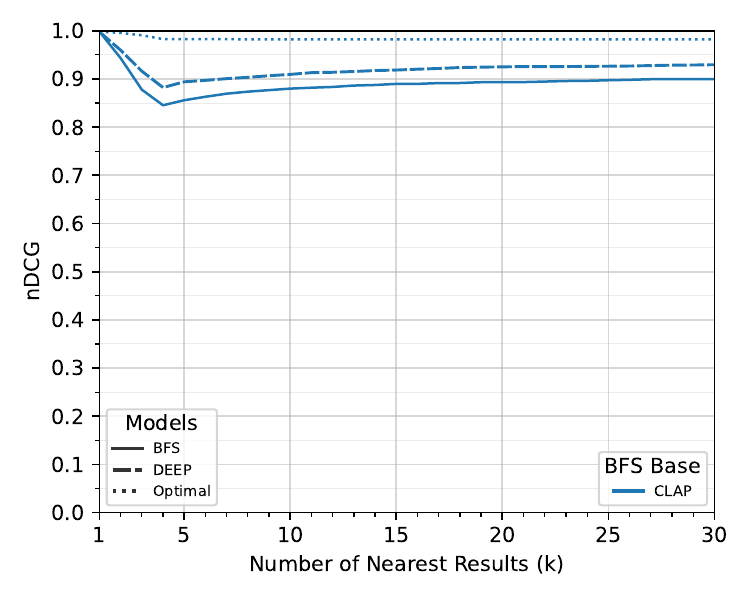}
    \caption{\ndcg for the \clapbfs model with \redeep re-ranking ($w=200$ and $k \in [1,30]$), averaged across 70 CVEs.}
    \label{img:VULN_ndcg_binpool}
\end{figure}

\begin{figure}[t!]
    \centering
    \includegraphics[width=\columnwidth, trim = 0cm 0cm 0cm 0cm]{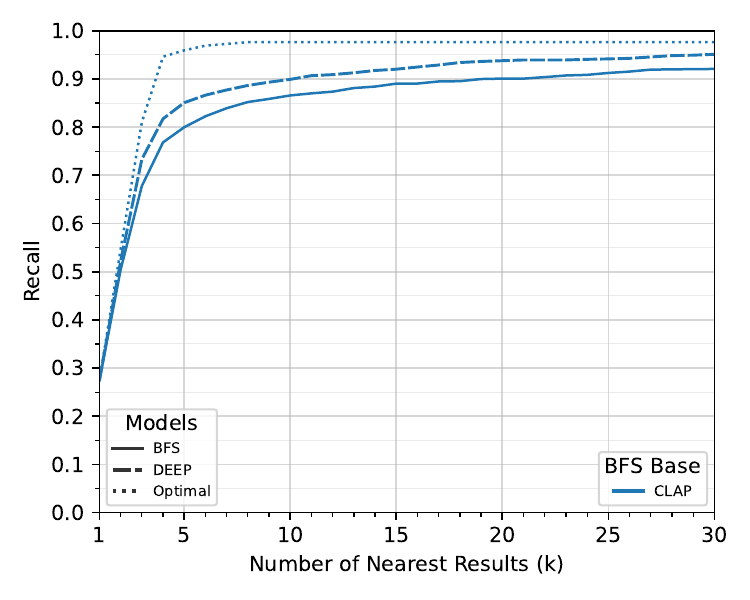}
    \caption{\recall for the \clapbfs model with \redeep re-ranking ($w=200$ and $k \in [1,30]$), averaged across 70 CVEs.}
    \label{img:VULN_recall_binpool}
\end{figure}

\section{Vulnerability Detection}\label{sec:vuln_det}

Following the approach of~\cite{jTrans-ISSTA22}, we evaluate the capability of \resafe of identifying vulnerable functions within packages affected by known CVEs. For this experiment, we consider CVEs extracted from the \binpool dataset~\cite{DBLP:conf/sigsoft/ArastehNWWPRH25}. This is a collection of real-world vulnerabilities extracted from different Debian packages and associated with specific CVEs, compiled for the x86-64 architecture using gcc with optimization levels ranging from O0 to O3. We follow the approach in Section~\ref{sec:impl_details} to disassemble the provided binaries; overall, our dataset comprises 138 CVEs affecting 64 distinct packages and 118 binaries across four Debian releases\footnote{bookworm, bullseye, buster, and stretch}, resulting in over 1.1 million functions.

We build an evaluation pool for each CVE, where each vulnerable function serves as a query against all functions from binaries affected by that CVE. We include only CVEs for which, after removing duplicates, we have at least 3 similar vulnerable functions. Figures~\ref{img:VULN_recall_binpool} and~\ref{img:VULN_ndcg_binpool} show the average Recall@$k$ and nDCG@$k$ across 70 CVE queries, with the dotted line indicating the performance of the optimal re-ranker.

The \clapbfs model achieves \ndcg of 0.90 and \recall of 0.92 at . When enhanced with \redeep, both metrics improve to 0.93 and 0.95 respectively, yielding a relative gain of +3\% for each metric and closing approximately half the gap to the optimal re-ranker. These improvements remain consistent across all $k$ values: \ndcg increases by +3.5\%, while \recall shows a gain of +6.2\% at $k=5$, closing roughly one-third of the gap to the optimum for both metrics at this rank. Note that this gap decreases monotonically as $k$ increases.

%% file: sections/related.tex

\section{Related Works}

This section surveys the main \bfs systems and re-ranking strategies in IR, NLP, and source-code retrieval.

\subsection{Binary Function Similarity}

Binary similarity research largely relies on embedding-based representations, where neural encoders map functions into dense semantic vectors. Apart from the embedding models used in this work, alternative approaches include \textbf{Asm2Vec}~\cite{ding2019asm2vec}, \textbf{Zeek}~\cite{shalev2018binary}, \textbf{CodeCMR}~\cite{DBLP:conf/nips/YuZW0NW20}, and \textbf{InnerEye}~\cite{DBLP:conf/ndss/ZuoLYL0Z19}, which capture features at the instruction level, represent code through intermediate program representations such as assembly or IR tokens, and encode local patterns of instruction sequences; \textbf{BinFinder}~\cite{10.1145/3579856.3582818} and \textbf{HermesSIM}~\cite{DBLP:conf/uss/HeLW0G0JWX24}, which improve robustness to obfuscation and compiler optimizations; \textbf{CI-Detector}~\cite{DBLP:conf/icse/Jia0XJW024} and \textbf{OrderMatters}~\cite{DBLP:conf/aaai/YuCTNHW20}, which explicitly use control-flow and cross-function dependencies; \textbf{PROTST}~\cite{DBLP:conf/saner/LuCLBC25} which uses a teacher-student approach where the transformer is trained on tasks of increasing difficulties. 

While all previous systems adopt embedding-based strategies, others adopt different approaches. Specifically, \textbf{GMN}~\cite{DBLP:conf/icml/LiGDVK19} is a GNN that implements a mechanism to match nodes across the CFGs of the functions to be analyzed; \textbf{Catalog1}~\cite{catalog} employs a fuzzy-hashing technique on raw bytes. Previous papers~\cite{marcelli2022machine,capozzi2023adversarial,artuso2022binbert}, have shown that these two models are inferior to some of the embedding models tested in our work. To the best of our knowledge, no existing function search system implements re-ranking.

\subsection{Re-ranking}

Re-ranking is a standard practice in information retrieval, aimed at refining an initial candidate set to improve ranking quality. Traditional learning-to-rank methods compute losses at the document, pair, or list level, while neural re-rankers extend this with representation-based (encode the query and candidate documents separately into embeddings), interaction-based (focus at the token or feature level interactions to capture rich contextual relationships), or hybrid architectures (combine both strategies to balance efficiency and fine-grained semantic understanding) \cite{DBLP:conf/cikm/HuangHGDAH13, DBLP:conf/cikm/ShenHGDM14, DBLP:conf/sigir/GangulyRMJ15, DBLP:journals/corr/MitraNCC16, DBLP:conf/cikm/GuoFAC16, DBLP:conf/www/Mitra0C17}. The advent of transformers \cite{devlin2018bert} and benchmarks such as \textbf{MS MARCO} \cite{DBLP:conf/nips/NguyenRSGTMD16} initiated neural re-ranking at scale \cite{lin2021pretrained}. 
Cross-encoder architectures capture contextual interactions efficiently, as demonstrated by \textbf{monoBERT} and \textbf{duoBERT} \cite{nogueira2019passage, nogueira2019multi}, while sequence-to-sequence models have been used for the generative-ranking paradigm estimating the likelyhood of a certain query given a certain document \cite{DBLP:conf/emnlp/NogueiraJPL20, DBLP:journals/corr/abs-2101-05667, DBLP:journals/corr/abs-1904-08375}. Hybrid re-ranking systems  combine several techniques in a complex pipeline, for example using non-neural ranking combined with a transformer-based ranking \cite{DBLP:journals/corr/abs-2004-13969}. 

The most related application of re-ranking is the one to high level source-code retrieval, demonstrating that multi-stage retrieval generalizes beyond natural language. For instance, some approaches combine BM25 with a neural re-ranker, while others refine fuzzy-matched candidates using sequential semantic similarity \cite{DBLP:journals/corr/abs-2502-07067, DBLP:journals/tosem/LiuXLLHL22}. These works show that applying a second-stage ranking is a recognized and effective practice, and our proposed method adheres to this established paradigm for binary function similarity.
We want to highlight that both text and high-level source code represent application settings that are widely different from the binaries used in our paper.

%% file: sections/conclusions.tex

\section{Limitations}\label{sec:limitations}
While our methodology presents significant advancements on the function search task and, broadly, on the binary function similarity problem, we identify certain limitations that could be addressed in future research. 

In our evaluation, we did not focus on temporal efficiency; future studies may involve exploring lightweight \reranker architectures and strategies. 

A natural future direction is cross-architecture generalization. Our experiments indicate that the \reranker generalizes across unseen x86-64 compiler toolchains; however, assessing its behavior on new architectures  (e.g., ARM64, MIPS) remains an open question. To further evaluate the generalizability of our approach, future research may involve the creation of a larger benchmark, including a broader set of instruction sets, a greater number of open-source projects, and binaries specific to practical use-cases of these systems (i.e., malware detection/classification, copyright infringement).

\section{Conclusions}
In this paper, we presented \resafe, a \reranker-based pipeline to enhance the performance of embedding-based BFS detection systems when employed in the function search task.

We demonstrated across two distinct datasets that our \reranker module, a decoder-only cross-encoder agnostic with respect to the underlying model, improves the ability of embedding-based BFS systems to retrieve, from large pools, functions compiled from the same source as a given query, in terms of \ndcg and \recall. 
We have shown quantifiable improvements in several scenario, closing the gap towards an optimal system.

%% file: sections/appendix.tex

\section{Impact on the Performance across Different Toolchains}\label{sec:app_B}

\begin{figure}[h!]
    \centering
    \includegraphics[width=\columnwidth, trim = 0cm 0cm 0cm 0cm]{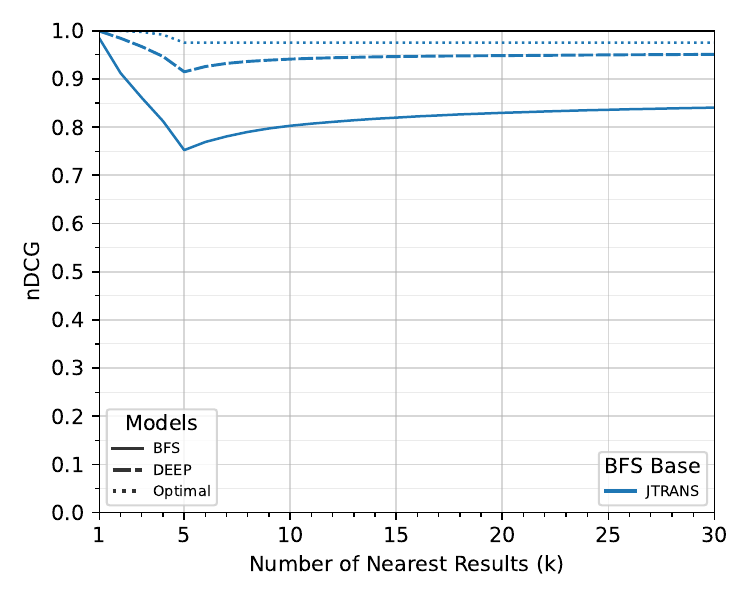}
    \caption{\ndcg for the \jtrans \bfs model using \redeep as the re-ranker, with $w=200$ and $K \in [1,30]$, evaluated on a pool of 25,000 functions and 5,000 queries from \bincorp.}
    \label{img:RQ2_jtrans_bincorp_1}
\end{figure}

\begin{figure}[h!]
    \centering
    \includegraphics[width=\columnwidth, trim = 0cm 0cm 0cm 0cm]{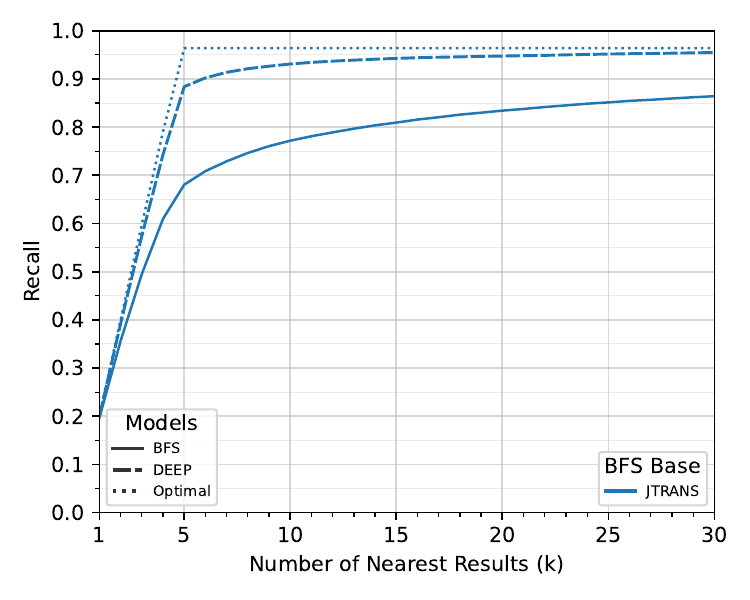}
    \caption{\recall for the \jtrans \bfs model using \redeep as the re-ranker, with $w=200$ and $K \in [1,30]$, evaluated on a pool of 25,000 functions and 5,000 queries from \bincorp.}
    \label{img:RQ2_jtrans_bincorp_2}
    
\end{figure}

\begin{figure}[h!]
    \centering
    \includegraphics[width=\columnwidth, trim = 0cm 0cm 0cm 0cm]{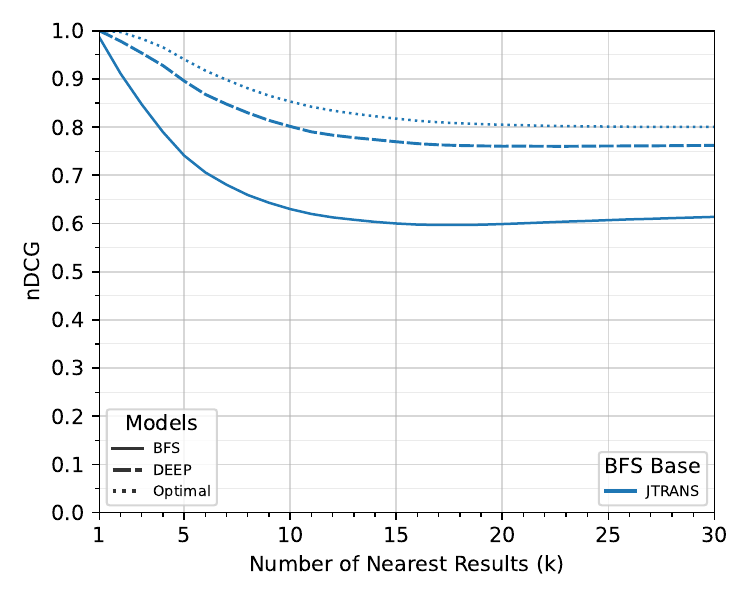}
    \caption{nDCG for the \jtrans \bfs model using \redeep as the \reranker, with $w=200$ and $K \in [1,30]$, evaluated on a pool of 11,622 functions and 1,000 queries from \multicomp.}
    \label{img:RQ2_jtrans_multicomp_1}
    
\end{figure}

\begin{figure}[h!]
    \centering
    \includegraphics[width=\columnwidth, trim = 0cm 0cm 0cm 0cm]{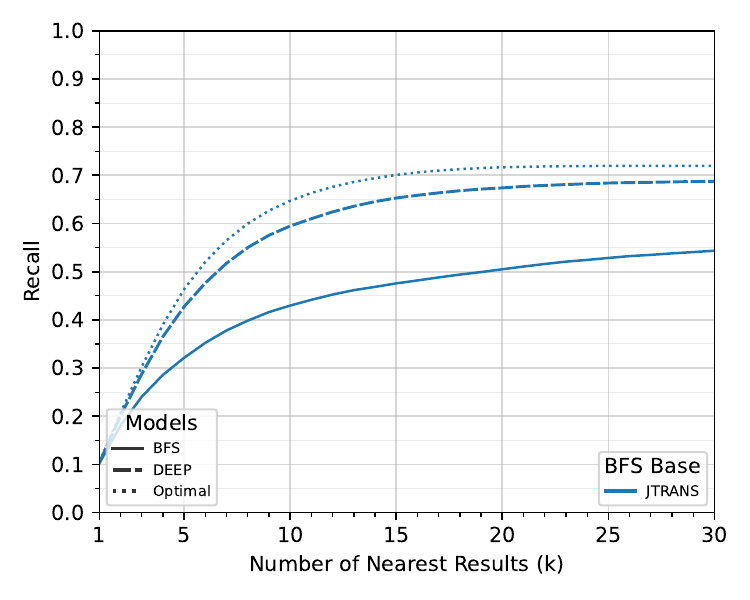}
    \caption{Recall for the \jtrans \bfs model using \redeep as the \reranker, with $w=200$ and $K \in [1,30]$, evaluated on a pool of 11,622 functions and 1,000 queries from \multicomp.}
    \label{img:RQ2_jtrans_multicomp_2}
    
\end{figure}

In the following, we report the results when re-ranking the results of \jtrans using our \redeep model. Figures~\ref{img:RQ2_jtrans_bincorp_1} and~\ref{img:RQ2_jtrans_bincorp_2} reports the \ndcg and \recall over the \bincorp dataset, while Figures~\ref{img:RQ2_jtrans_multicomp_1} and~\ref{img:RQ2_jtrans_multicomp_2} reports the results over \multicomp.

\clearpage

\begin{table*}[h!]
\centering
\caption{\ndcg and \recall for the considered \bfs models using \redeep as the \reranker, with $w \in \{30, 50, 100, 200\}$ and $k \in \{5, 10, 15, 20, 25, 30\}$, evaluated on the \bincorp and \multicomp datasets.
} 
\label{tab:RQ2}
\resizebox{\textwidth}{!}{%
\begin{tabular}{l|l|c|l|rrrrrr|rrrrrr}
\cmidrule[\heavyrulewidth]{5-16}
\multicolumn{4}{l}{} & \multicolumn{6}{c|}{\textbf{\ndcg}} & \multicolumn{6}{c}{\textbf{\recall}} \\
\midrule
\textbf{Dataset} & \textbf{\bfs model} & \textbf{Window size ($w$)} & \textbf{\reranker} &
\textbf{@5} & \textbf{@10} & \textbf{@15} & \textbf{@20} & \textbf{@25} & \textbf{@30} &
\textbf{@5} & \textbf{@10} & \textbf{@15} & \textbf{@20} & \textbf{@25} & \textbf{@30} \\
\cmidrule{1-16}

\multirow{15}{*}{\bincorp}
 & \multirow{5}{*}{\jtrans} & $-$ & \xmark & 0.75 & 0.80 & 0.82 & 0.83 & 0.84 & 0.84 & 0.68 & 0.77 & 0.81 & 0.83 & 0.85 & 0.86 \\
 &  & 30 & \cmark & 0.88 & 0.89 & 0.90 & 0.90 & 0.90 & 0.90 & 0.84 & 0.86 & 0.86 & 0.86 & 0.86 & 0.86 \\
 &  & 50 & \cmark & 0.90 & 0.91 & 0.91 & 0.92 & 0.92 & 0.92 & 0.86 & 0.89 & 0.89 & 0.89 & 0.90 & 0.90 \\
 &  & 100 & \cmark & 0.91 & 0.93 & 0.94 & 0.94 & 0.94 & 0.94 & 0.88 & 0.92 & 0.92 & 0.93 & 0.93 & 0.93 \\
 &  & 200 & \cmark & 0.91 & 0.94 & 0.95 & 0.95 & 0.95 & 0.95 & 0.88 & 0.93 & 0.94 & 0.95 & 0.95 & 0.95 \\
\cmidrule{2-16}
 & \multirow{5}{*}{\clapbfs} & $-$ & \xmark & 0.87 & 0.89 & 0.90 & 0.91 & 0.91 & 0.91 & 0.83 & 0.87 & 0.89 & 0.90 & 0.91 & 0.91 \\
 &  & 30 & \cmark & 0.91 & 0.93 & 0.93 & 0.93 & 0.93 & 0.93 & 0.88 & 0.91 & 0.91 & 0.91 & 0.91 & 0.91 \\
 &  & 50 & \cmark & 0.91 & 0.93 & 0.93 & 0.94 & 0.94 & 0.94 & 0.88 & 0.91 & 0.92 & 0.92 & 0.93 & 0.93 \\
 &  & 100 & \cmark & 0.91 & 0.93 & 0.94 & 0.94 & 0.94 & 0.94 & 0.88 & 0.92 & 0.93 & 0.93 & 0.94 & 0.94 \\
 &  & 200 & \cmark & 0.91 & 0.94 & 0.94 & 0.94 & 0.94 & 0.94 & 0.88 & 0.92 & 0.94 & 0.94 & 0.94 & 0.95 \\
\cmidrule{2-16}
 & \multirow{5}{*}{\binbert} & $-$ & \xmark & 0.74 & 0.78 & 0.79 & 0.80 & 0.80 & 0.81 & 0.66 & 0.73 & 0.76 & 0.78 & 0.79 & 0.80 \\
 &  & 30 & \cmark & 0.84 & 0.85 & 0.85 & 0.85 & 0.85 & 0.85 & 0.78 & 0.80 & 0.80 & 0.80 & 0.80 & 0.80 \\
 &  & 50 & \cmark & 0.85 & 0.86 & 0.87 & 0.87 & 0.87 & 0.87 & 0.80 & 0.82 & 0.82 & 0.82 & 0.83 & 0.83 \\
 &  & 100 & \cmark & 0.87 & 0.88 & 0.89 & 0.89 & 0.89 & 0.89 & 0.82 & 0.85 & 0.85 & 0.86 & 0.86 & 0.86 \\
 &  & 200 & \cmark & 0.88 & 0.90 & 0.90 & 0.91 & 0.91 & 0.91 & 0.84 & 0.87 & 0.88 & 0.88 & 0.88 & 0.89 \\

\midrule
\midrule

\multirow{15}{*}{\multicomp}
 & \multirow{5}{*}{\jtrans} & $-$ & \xmark & 0.74 & 0.63 & 0.60 & 0.60 & 0.61 & 0.61 & 0.32 & 0.43 & 0.48 & 0.50 & 0.53 & 0.54 \\
 &  & 30 & \cmark & 0.84 & 0.72 & 0.67 & 0.66 & 0.65 & 0.65 & 0.39 & 0.50 & 0.53 & 0.54 & 0.54 & 0.54 \\
 &  & 50 & \cmark & 0.86 & 0.74 & 0.70 & 0.69 & 0.68 & 0.68 & 0.40 & 0.53 & 0.56 & 0.58 & 0.58 & 0.58 \\
 &  & 100 & \cmark & 0.88 & 0.77 & 0.74 & 0.73 & 0.72 & 0.73 & 0.41 & 0.56 & 0.61 & 0.63 & 0.63 & 0.64 \\
 &  & 200 & \cmark & 0.90 & 0.80 & 0.77 & 0.76 & 0.76 & 0.76 & 0.43 & 0.59 & 0.65 & 0.67 & 0.68 & 0.69 \\
\cmidrule{2-16}
 & \multirow{5}{*}{\clapbfs} & $-$ & \xmark & 0.88 & 0.81 & 0.79 & 0.80 & 0.80 & 0.81 & 0.42 & 0.62 & 0.70 & 0.74 & 0.77 & 0.78 \\
 &  & 30 & \cmark & 0.92 & 0.85 & 0.83 & 0.82 & 0.83 & 0.83 & 0.45 & 0.66 & 0.73 & 0.76 & 0.78 & 0.78 \\
 &  & 50 & \cmark & 0.92 & 0.86 & 0.84 & 0.84 & 0.84 & 0.84 & 0.45 & 0.67 & 0.75 & 0.78 & 0.79 & 0.80 \\
 &  & 100 & \cmark & 0.92 & 0.86 & 0.84 & 0.84 & 0.85 & 0.85 & 0.45 & 0.67 & 0.76 & 0.79 & 0.81 & 0.82 \\
 &  & 200 & \cmark & 0.92 & 0.86 & 0.84 & 0.84 & 0.85 & 0.85 & 0.45 & 0.67 & 0.76 & 0.80 & 0.82 & 0.83 \\
\cmidrule{2-16}
 & \multirow{5}{*}{\binbert} & $-$ & \xmark & 0.91 & 0.86 & 0.84 & 0.84 & 0.85 & 0.86 & 0.45 & 0.67 & 0.76 & 0.80 & 0.83 & 0.84 \\
 &  & 30 & \cmark & 0.95 & 0.90 & 0.88 & 0.88 & 0.88 & 0.88 & 0.47 & 0.72 & 0.80 & 0.83 & 0.84 & 0.84 \\
 &  & 50 & \cmark & 0.95 & 0.91 & 0.89 & 0.89 & 0.89 & 0.89 & 0.47 & 0.73 & 0.82 & 0.85 & 0.87 & 0.87 \\
 &  & 100 & \cmark & 0.95 & 0.91 & 0.90 & 0.90 & 0.90 & 0.91 & 0.47 & 0.73 & 0.82 & 0.86 & 0.88 & 0.89 \\
 &  & 200 & \cmark & 0.94 & 0.91 & 0.90 & 0.90 & 0.90 & 0.91 & 0.47 & 0.72 & 0.82 & 0.87 & 0.88 & 0.89 \\

\bottomrule
\end{tabular}
}
\end{table*}

\section{Impact of the Window Size}\label{sec:app_A}

Table~\ref{tab:RQ2} shows the impact of the window size $w$ on \resafe. 

Overall, $w=200$ yields the best performance across all the models and pools. However, depending on the considered \bfs model and the considered pool \pool, $w$ differently influences the performance. Considering \bincorp, there increasing $w$ constantly improves the performance when re-ranking results from \jtrans and \binbert. However, this is not the case of \clapbfs: for $k \le 10$, the improvement is below 1\% in average; thus, while reducing $w$ may not affect the overall performance, this can produce a benefit from a computational point of view. 

Moving to \multicomp, \jtrans is the model that is mostly affected by the $w$ parameter, with a monotonically increasing increment across the considered values of $k$; contrarily, when considering \clapbfs and \binbert with low values for $k$ (i.e., $k \le 10$), increasing $w$ does not yield to a substantial improvement.

\begin{table*}[t!]
\centering
\caption{\ndcg and \recall scores when considering $w=200$ and $k \in \{5, 10, 15, 20, 25, 30\}$, evaluated on a pool of 25,000 functions extracted from the \bincorp dataset. The \ensemble row shows the performance obtained when ensembling the re-ranked results from \jtrans and \binbert.}
\scriptsize
\label{tab:bincorp_ensemble}
\begin{tabular}{l|l|l|rrrrrr|rrrrrr}
\cmidrule{4-15}
\multicolumn{3}{l}{} & \multicolumn{6}{c|}{\textbf{nDCG}} & \multicolumn{6}{c}{\textbf{Recall}} \\
\midrule
\textbf{Dataset} & \textbf{\bfs model} & \textbf{\reranker} & \textbf{@5} & \textbf{@10} & \textbf{@15} & \textbf{@20} & \textbf{@25} & \textbf{@30} & \textbf{@5} & \textbf{@10} & \textbf{@15} & \textbf{@20} & \textbf{@25} & \textbf{@30} \\
\midrule
\multirow{7}{*}{\bincorp} 
& \multirow{2}{*}{\jtrans} & \xmark & 0.75 & 0.80 & 0.82 & 0.83 & 0.84 & 0.84 & 0.68 & 0.77 & 0.81 & 0.83 & 0.85 & 0.86 \\
&                       & \cmark & 0.91 & 0.94 & 0.95 & 0.95 & 0.95 & 0.95 & 0.88 & 0.93 & 0.94 & 0.95 & 0.95 & 0.95 \\
\cmidrule{2-15}
& \multirow{2}{*}{\clapbfs} & \xmark & 0.87 & 0.89 & 0.90 & 0.91 & 0.91 & 0.91 & 0.83 & 0.87 & 0.89 & 0.90 & 0.91 & 0.91 \\
&                       & \cmark & 0.91 & 0.94 & 0.94 & 0.94 & 0.94 & 0.94 & 0.88 & 0.92 & 0.94 & 0.94 & 0.94 & 0.95 \\
\cmidrule{2-15}
& \multirow{2}{*}{\binbert} & \xmark & 0.74 & 0.78 & 0.79 & 0.80 & 0.80 & 0.81 & 0.66 & 0.73 & 0.76 & 0.78 & 0.79 & 0.80 \\
&                       & \cmark & 0.88 & 0.90 & 0.90 & 0.91 & 0.91 & 0.91 & 0.84 & 0.87 & 0.88 & 0.88 & 0.88 & 0.89 \\
\cmidrule{2-15}
& \multicolumn{2}{c|}{\ensemble} & 0.92 & 0.95 & 0.95 & 0.96 & 0.96 & 0.96 & 0.89 & 0.94 & 0.95 & 0.96 & 0.97 & 0.97 \\
\midrule
\midrule
\multirow{7}{*}{\multicomp} 
& \multirow{2}{*}{\jtrans} & \xmark & 0.74 & 0.63 & 0.60 & 0.60 & 0.61 & 0.61 & 0.32 & 0.43 & 0.48 & 0.50 & 0.53 & 0.54 \\
&                       & \cmark & 0.90 & 0.80 & 0.77 & 0.76 & 0.76 & 0.76 & 0.43 & 0.59 & 0.65 & 0.67 & 0.68 & 0.69 \\
\cmidrule{2-15}
& \multirow{2}{*}{\clapbfs} & \xmark & 0.88 & 0.81 & 0.79 & 0.80 & 0.80 & 0.81 & 0.42 & 0.62 & 0.70 & 0.74 & 0.77 & 0.78 \\
&                       & \cmark & 0.92 & 0.86 & 0.84 & 0.84 & 0.85 & 0.85 & 0.45 & 0.67 & 0.76 & 0.80 & 0.82 & 0.83 \\
\cmidrule{2-15}
& \multirow{2}{*}{\binbert} & \xmark & 0.91 & 0.86 & 0.84 & 0.84 & 0.85 & 0.86 & 0.45 & 0.67 & 0.76 & 0.80 & 0.83 & 0.84 \\
&                       & \cmark & 0.94 & 0.91 & 0.90 & 0.90 & 0.90 & 0.91 & 0.47 & 0.72 & 0.82 & 0.87 & 0.88 & 0.89 \\
\cmidrule{2-15}
& \multicolumn{2}{c|}{\ensemble} & 0.94 & 0.90 & 0.89 & 0.90 & 0.90 & 0.91 & 0.47 & 0.72 & 0.82 & 0.87 & 0.89 & 0.90 \\
\bottomrule
\end{tabular}
\end{table*}

\section{Ensembling Performance on \bincorp and \multicomp}\label{sec:app_C}

In the following, we analyze the impact of our ensembling strategy on pools extracted from \bincorp and \multicomp (see Table~\ref{tab:bincorp_ensemble} for complete results). Details on these pools are provided in Section~\ref{sec:eval_dataset}.

On \bincorp, our ensemble outperforms the strongest individual re-ranked model (\redeep on \jtrans) by 1\% in \ndcg and 2\% in \recall on average. On \multicomp, it matches the performance of re-ranked \binbert across all $k$ values, consistently ranking \binbert's top results at the highest positions.